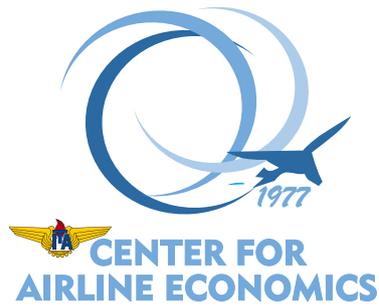

# WORKING PAPER SERIES

Airport service quality perception and flight delays: examining the influence of psychosituational latent traits of respondents in passenger satisfaction surveys

Alessandro V. M. Oliveira
Bruno F. Oliveira
Moisés D. Vassallo

# Airport service quality perception and flight delays: examining the influence of psychosituational latent traits of respondents in passenger satisfaction surveys

Alessandro V. M. Oliveira✈
Bruno F. Oliveira
Moisés D. Vassallo

This version: 10 September 2023

## Abstract

The service quality of a passenger transport operator can be measured through face-to-face surveys at the terminals or on board. However, the resulting responses may suffer from the influence of the intrinsic aspects of the respondent's personality and emotional context at the time of the interview. This study proposes a methodology to generate and select control variables for these latent psychosituational traits, thus mitigating the risk of omitted variable bias. We developed an econometric model of the determinants of passenger satisfaction in a survey conducted at the largest airport in Latin America, São Paulo GRU Airport. Our focus was on the role of flight delays in the perception of quality. The results of this study confirm the existence of a relationship between flight delays and the global satisfaction of passengers with airports. In addition, favorable evaluations regarding airports' food/beverage concessions and Wi-Fi services, but not their retail options, have a relevant moderating effect on that relationship. Furthermore, dissatisfaction arising from passengers' interaction with the airline can have negative spillover effects on their satisfaction with the airport. We also found evidence of blame-attribution behavior, in which only delays of internal origin, such as failures in flight management, are significant, indicating that passengers overlook weather-related flight delays. Finally, the results suggest that an empirical specification that does not consider the latent psychosituational traits of passengers produces a relevant overestimation of the absolute effect of flight delays on passenger satisfaction.

*Keywords*: transport, survey, psychology, satisfaction, delay.

JEL Classification: D22; L11; L93.

✈Corresponding author. Email address: alessandro@ita.br.

▪ Affiliations: Center for Airline Economics, Aeronautics Institute of Technology, Brazil (first two authors); Federal University of Itajubá, Brazil (third author).

▪ Acknowledgments: The first author wishes to thank the São Paulo Research Foundation (FAPESP), grant n. 2020-06851, the National Council for Scientific and Technological Development (CNPq), grant numbers 301344/2017-5 and 305439/2021-9. The second author thanks the Coordination for the Improvement of Higher Education Personnel (CAPES), Finance Code 001. The authors are grateful to Ewerton Monti, Tatiana Repetto, Thiago Greghi, Luiz Gustavo Souza, Mauro Caetano, Marcelo Guterres, Evandro Silva, Giovanna Ronzani, Rogéria Arantes, Cláudio Jorge Pinto Alves, Gui Lohmann, Wallace Souza, Maria Clara Ramos. All mistakes are ours.

# 1. Introduction

The post-pandemic world has been characterized by strong pressure on logistics competitiveness to control transport costs and prices. In this context, effective management of the quality of services provided by transport operators has become increasingly crucial to conquer market advantage and survive in such an ever-changing market environment. Furthermore, as critical as quality is, the management of customers' perceptions is also key to transport operators, which must be scrutinized before, during, and after service provision. In the Business-to-Consumer (B2C) segment, monitoring transport customer feedback on social media is essential for fostering profitability and brand recognition.

In today's aviation industry, the quality of airport services plays a crucial role in shaping passengers' overall experience. Airports, as the first and last point of contact in air travel, are fundamental in shaping passengers' perceptions and views as consumers. The quality of services, including check-in procedures, security screening, cleanliness, amenities, and staff efficiency, significantly influences passengers' perception of an airport. On the other hand, passengers' emotional context, such as their expectations, anxieties, and perspectives, which may vary based on annual, weekly, and even daily seasonality, can also impact these perceptions and the formation of their overall satisfaction. In this regard, flight delays can have a profound impact on the perceived quality of airport services. Delays disrupt travel schedules, cause inconvenience, and can lead to frustration and dissatisfaction among passengers. The longer the delays, the greater the likelihood of passengers forming negative opinions about the quality of an airport's services. Passengers may associate delays with poor management, lack of infrastructure, or ineffective communication, resulting in a diminished perception of the overall airport quality and potentially leading to formal complaints to the airport administration and negative experiences shared on social media. Therefore, addressing and minimizing flight delays is essential for airports to maintain a positive reputation, improve customer satisfaction, and ensure continued growth and competitiveness in the business.

In the air travel industry, service quality is commonly measured through passenger assessments through face-to-face surveys conducted on boards or terminals. Respondents rate their overall impressions of service quality ("global satisfaction") by choosing alternatives on a Likert scale. They may also be invited to rate several aspects of the provision of services with respect to service attributes, stages, and available facilities ("local satisfaction"). Typically, the survey also contains questions on the socioeconomic and behavioral characteristics of the respondents so that operators can gather additional information about the consumer profile.

One of the challenges for researchers who conduct empirical studies using databases derived from surveys is the consideration of unobserved determinants of satisfaction. In particular, it can be argued that the collected responses may be influenced by various psychological traits of the respondents, which are usually not the target of satisfaction surveys. These psychological traits refer to aspects of personality and issues that emerge during the interview. For example, an individual with a more optimistic view may rate the service above the respondent's mean. However, an individual who experiences a specific problem that generates impatience or irritability may rate the same service below the mean. If factors such as these are purely random, the analyses generated from the survey data will not present major problems.

However, part of the psychological load of the survey respondents can be formed in a way that is not necessarily random. For example, periods of demand peak for the operator's services may also be moments of greater impatience caused by other factors, such as urban traffic jams or public transport overcrowding experienced before arrival at the airport. In this case, satisfaction may be correlated with confounding psychological factors of passengers, which makes empirical analyses that use survey datasets vulnerable to omitted variable bias.

This study proposes a methodology for generating and selecting control variables to account for the latent psychosituational traits of survey respondents in empirical models of passenger behavior and attitude. Our framework is grounded in elements of the Motivation Research, and the Happiness Economics literature, which delve into individuals' satisfaction across various aspects of life, and hold implications for traveler psychology and consumption behavior. Specifically, we draw upon the meta-theoretical framework of Mowen (2000) and the research conducted by van Praag and Ferrer-i-Carbonell (2008).

We developed an econometric model of the determinants of passenger satisfaction based on a survey conducted at the largest airport in Latin America, São Paulo/Guarulhos Airport (GRU Airport). As a case study, we focus on an empirical analysis of the relationship between flight delay and perceived quality. Because our proposal considers a broad set of controls, we employ a High-Dimensional Sparse Regression method for model selection. We utilize the Least Absolute Shrinkage and Selection Operator (LASSO) procedure of Belloni et al. (2012), Belloni, Chernozhukov, and Hansen (2014a, b), and Chernozhukov, Hansen, and Spindler (2015).

We aim to contribute by developing an extension of the satisfaction model to consider the possibility of passengers attributing blame (Anderson, Bagget and Widener, 2009). We develop a two-step probit/LASSO-based model to distinguish between delays of "internal" origin, i.e., stemming from operational issues such as flight management failures and runway congestion, and delays of "external" origin, i.e., due to unfavorable weather. Our approach allows empirical testing and checking the incidence of both types of delays on the satisfaction formation of passengers.

Our study integrates with and adds value to the existing literature by confirming and expanding upon previous findings. We examine the relationship between flight delays and consumer satisfaction and well-being in airports, as accomplished by many previous studies, such as Anderson, Bagget and Widener (2009), Britto, Dresner, and Voltes (2012), Song, Guo, and Zhuang (2020), and Wen and Geng-qing Chi (2013). We align with Anderson, Bagget and Widener (2009) in acknowledging the critical role of blame attribution in shaping customer satisfaction during service failures in the airline industry. Additionally, our research agrees with Britto, Dresner, and Voltes (2012) in highlighting the correlation between flight delays and consumer well-being. Moreover, our study aligns with Wen and Geng-qing Chi (2013) in recognizing the influence of customer emotions on service evaluations. We contribute by utilizing specific control variables as proxies for the psychosituational latent traits of respondents in an airport passenger satisfaction survey in Brazil. Additionally, our investigation concentrates on overall passenger satisfaction with airports while considering the moderating effects of specific airport services, an element that is scarcely investigated by the previous literature.

The remainder of this paper is organized as follows. Section 2 discusses the existing literature on survey respondents' personality and emotional context and the relationship between service failure and the attribution of blame. Section 3 introduces the research design, with a description of GRU Airport, the available data, and our proposed econometric model. Section 4 outlines our empirical methodology. Section 5 presents the estimation results for the airport satisfaction model. Finally, Section 6 delivers the study's conclusions.

## 2. Literature review

### 2.1. Passenger satisfaction and the assessment of airport service quality

Air transport service quality is a well-explored topic in academic literature, with granular studies analyzing each aspect of this theme in the industry. For example, studies from Bellizzi, Eboli & Mazzulla (2020) and Eboli, Bellizzi & Mazzulla (2022) show that this vast literature can be analyzed from the point of view passengers, and further, separated according to the type of service in the industry (e.g., airport and airline service quality), compiling the main findings in this field in the last decade (2008-2020), how studies collected passengers data and data analysis methods. Because of the dynamic nature of the air transport field, there are even recent research that adapts its focus in response to the unprecedented challenges of the COVID-19 pandemic. Usman et al. (2022) conduct a literature review outlining the evolution of airport service quality (ASQ) research and highlight the need for further investigations to reconcile general service quality measures with those specific to the airport industry. Rocha, Costa, and Silva (2022) perform a comprehensive review and bibliometric analysis on quality of service (QoS) evaluations at airport terminals, identifying key areas lacking specific studies and suggesting the inclusion of sustainability and COVID-19 related criteria in future evaluations.

Notably, several recent studies apply data-driven methodologies to these questions. Lopez-Valpuesta and Casas-Albala (2023) utilize ASQ surveys and econometric modeling to examine how passenger satisfaction at Seville Airport in Spain is influenced by a variety of factors, such as passenger nationality, travel motives, destination, and airport cleanliness and comfort. Li et al. (2022) implement a data-driven crowdsourcing approach, analyzing Google Maps reviews from the 98 busiest U.S. airports through sentiment analysis, noting more positive sentiments towards the environment and personnel, yet unchanged views on facilities. Meanwhile, Burrieza-Galán et al. (2022) develop a novel methodology using anonymized mobile phone records and airport surveys to analyze passenger behavior and improve service quality.

Numerous studies have been conducted to assess and understand the quality of airport services. Chonsalasin, Jomnonkwao, and Ratanavaraha (2021) develop a measurement model to determine passengers' expectations of service quality at airports in Thailand. Prentice and Kadan (2019) investigate the impact of ASQ on both airport and destination choice, highlighting the relationship between service quality, passenger satisfaction, and behavioral intentions. Halpern and Mwesiumo (2021) examine the consequences of service failures on passenger satisfaction and their inclination to promote an airport online. Barakat, Yeniterzi, and Martin-Domingo (2021) utilize deep learning models to analyze Twitter data and measure ASQ, focusing on sentiment analysis. Antwi et al. (2020) explore the influence of processing and non-processing service quality on passenger satisfaction and affective image, emphasizing the significance of service attributes in shaping passengers' perceptions. Martin-Domingo, Martín, and Mandsberg (2019) employ social media as a resource for sentiment analysis, providing insights into the aspects of ASQ valued and criticized by passengers. According to their sentiment analysis, London Heathrow Airport excels in terms of Wi-Fi availability, restroom facilities, food and beverage options, and lounge services. However, areas such as waiting areas, parking facilities, passport arrival procedures, staff service, and passport control are identified as areas that require improvement. Pagliari and Graham (2019) analyze the effects of ownership change on airport competition, uncovering dynamics related to pricing strategies, route development, and quality of service. Lastly, Cao, Li, and Zhang (2023) present a case study emphasizing the relevance of evaluating service satisfaction and prioritizing areas for improvement from the passengers' perspective.

Furthermore, the systemic implications of flight delays on the interconnected network of airlines, airports, and tourism destinations have been investigated, uncovering their significant impact on the perception of ASQ. Efthymiou, Arvanitis and Papatheodorou (2016) explore institutional changes in the European aviation sector and highlight the positive effects of adopting a systemic approach and relaxing regulatory and infrastructural constraints on tourism growth in both central and peripheral regions. Additionally, Efthymiou et al. (2019) examine British Airways' on-time performance at London Heathrow Airport, revealing that the airline consistently meets or exceeds customer expectations, contrary to the belief that passengers are frustrated by delays. These studies emphasize the ability of airlines to satisfy customers and retain their loyalty.

Another aspect of flight delays that recent literature addresses is the relationship between flight delays and customer choice and demand. Several articles shed light on this topic. Britto, Dresner, and Voltes (2012) examine the impact of flight delays on passenger demand and airfares. Their findings indicate that flight delays decrease passenger demand and lead to higher airfares, resulting in significant decreases in both consumer and producer welfare. Arora and Mathur (2020) investigate the effect of airline choice and temporality on flight delays, identifying consumer-centric factors as predictors of departure and arrival delays. Gayle and Yimga (2018) find robust evidence of consumers valuing air travel on-time performance and being willing to pay to avoid delays. Song, Guo, and Zhuang (2020) explore the emotions expressed by passengers in response to flight delays, based on an analysis of comments from the SKYTRAX platform, with potential implications on passenger choice. Yimga (2020) investigates the impact of on-time performance on price and marginal cost, revealing that flight delays have negative effects on consumer welfare. Jiang and Ren (2019) investigate the decision-making behavior of passengers when faced with flight delays, taking into account the dynamic and heterogeneous nature of their reference point. Finally, Yimga and Gorjidooz (2019) emphasize the significance of on-time performance (OTP) for consumers and the public disclosure of airline OTP, which create incentives for carriers to extend their scheduled gate-to-gate times and engage in aggressive schedule padding. By employing a discrete choice demand model for air travel, they investigate the impact of schedule padding on consumer preferences and decision-making.

While the aforementioned studies have greatly advanced our comprehension of airport service quality, our research takes on a distinct approach by centering on the impact of passenger personality traits and emotional context in shaping their responses to ASQ surveys. We contend that these elements can significantly influence passengers' perceptions of their experiences, subsequently affecting their reported satisfaction levels. In the following sections, we delve into the related literature concerning this topic.

## 2.2. Personality and emotional context of survey respondents

Customer satisfaction is a recognized antecedent of brand loyalty (Noyan and Şimşek, 2014) and is one of the most important goals in a company's quality management. It is reasonable to argue that dissatisfied customers can negatively impact the firm's profitability and value (Anderson, Fornell and Mazvancheryl, 2004; Hult et al., 2019).

In transportation, carriers and terminal operators use service quality surveys to understand, monitor, and influence passenger satisfaction. These surveys, whether conducted face-to-face at terminals or onboard, or online during the post-travel period, are generally configured with Likert-scale questions in which the respondents rate their perception of overall quality service (global satisfaction) and their perception of quality concerning the stages of the service provided and their associated facilities (local satisfaction). Examples of recent studies that analyze passenger survey data include Laisak, Rosli, and Sa'adi (2021), He, Yang, and Li (2021) for buses, Monsuur et al. (2021) for trains, Yuan et al. (2021) for the integration between air-rail transport, Halpern and Mwesiumo (2021), Mainardes, Melo and Moreira (2021), and Bezerra and Gomes (2020) for airports, among many others.[1]

As assessed by survey responses, one of the most important drivers of satisfaction is related to the respondent's personality and situational and emotional contexts. Personality is usually defined in the literature as a set of psychological characteristics of an individual that explain lasting and distinct patterns of feelings, thoughts, and behaviors (Pervin and Cervone, 2019; Smith, 2020). Studies such as Schul and Crompton (1983), Mowen (2000), and Smith (2020) highlight that personality characteristics are responsible for a relevant part of the variation in consumer behavior and are more effective than demographic variables in predicting purchasing behavior. One of the reasons for such performance is that personality traits are generally very stable and non-transient compared to other types of individual attributes, such as mood, attitudes, and income.

Mowen (2000) used the trait approach in his meta-theoretical model of motivation and personality. The author's object of study centers on how personality interacts with situational context to influence individuals' feelings, thoughts, and behavior. The conceptual framework of the author assumes that a set of "elemental traits" of the individual – that is, those characteristics arising from their genetic idiosyncrasies and early learning – combine with the environment to create "compound traits," which combine with the situation to form "situational traits." These situational traits would be "*predispositions to act within general contexts of behavior.*" [2] Additionally, these situational traits can combine with compound and elemental traits, resulting in "surface traits," i.e., "*enduring tendencies to act with respect to specific categories of behavior.*"[2]

---

[1] There are also many passenger satisfaction surveys conducted by transport authorities and institutions around the world. For example, in air transport, notable examples are the surveys conducted by global aviation organizations such as the Airport Service Quality (ASQ) of the Airports Council International (ACI), and the Global Passenger Survey (GPS), of the International Air Transport Association (IATA). Other examples are the Bus Passenger Survey, the National Rail Passenger Survey (Transport Focus, UK), and the Departing Passenger Survey (CAA, UK), and Transport for NSW's surveys for the Customer Satisfaction Index report (New South Wales, Australia). There are also surveys conducted by the operators themselves, such as the Annual Customer Satisfaction Survey at San Francisco Airport, in the United States, and the Passenger Satisfaction Survey at Amsterdam Schiphol Airport, Netherlands, among many others.

[2] Mowen (2000), p. 6.

Regarding the situational context of the individual, some studies have discussed the characteristics of information processing by humans, which their emotions can influence. For example, anxious and depressed people tend to judge negatively about past, present, or future events and interpret ambiguous stimuli unfavorably (Harding, Paul and Mendl 2004). Iversen, Kupfermann, and Kandel (2000) consider that an emotional state contains two components: characteristic physical sensation (for example, feeling one's heartbeat) and conscious feeling (for example, consciously feeling afraid).

Wright and Bower (1992) found that a person's mood can affect a judgment about the uncertainty of a future event. The authors investigate subjective probabilities reported by individuals classified in a "happy," "neutral," or "sad" mood. Their results point to the presence of relevant and intuitive mood effects, indicating that happy people are "optimistic," that is, they report greater probabilities of positive future events and lower probabilities of negative events. On the other hand, sad people are "pessimistic," providing lower probabilities for positive events and higher probabilities for negative events. MacLeod and Byrne (1996) also study the effects of anticipating positive and negative future experiences on mood states. Their results suggest that anxious individuals anticipate negative future experiences more, while anxiety-depressed individuals present greater anticipation of negative experiences and lower anticipation of positive experiences.

*2.3. Service failure and attribution of blame*

One of the relevant psychological issues related to the association between satisfaction with the provision of the service and the additional wait caused by its failure concerns the attribution of blame. If customers blame the service provider for the delay, then in principle, the reported overall satisfaction could be even more impaired.

The literature on the relationship between service failure and the attribution of blame is vast. Harris et al. (2006) examined differences in consumers' attributions of blame for service failures and their effects on their recovery expectations in online and offline environments. Forrester and Maute (2001) find that relationship-building efforts by firms reduce the likelihood and intensity of adverse customer behaviors, such as less intense blame and anger. Their results suggest strategies to reduce vulnerability to customer defection and avoid adverse communication, such as negative word of mouth.

Nikbin et al. (2011) assessed the impact of company reputation on customer responses to service failure by focusing on the effect of blame attribution. The authors detected the moderating role of failure attributions in the relationship between company reputation and behavioral intentions. Choi and Mattila (2008) studied the impact of perceived controllability on service failures and service quality expectations on customer reactions to these failures. Their results indicated that customers react negatively when they believe that the service provider can easily avoid failure. Additionally, when customers feel partially responsible for a failure, the negative effects of poor performance are mitigated. They also point out that high service quality expectations mitigate negative effects. Weber and Sparks (2004) studied consumer attributions and behavioral responses to service failures in airlines' strategic alliance configurations. The authors pointed to a negative spillover effect on an airline in an alliance arising from the service failure of a partner airline. They indicate that these effects are a negative evaluation, dissatisfaction, negative word-of-mouth, and reduced loyalty.

Anderson, Bagget and Widener (2009) investigated the impact of flight delays on customer satisfaction in the US airline industry. They use passenger survey data obtained from a market research firm. The authors examined the moderating role of blame attribution on the effects of service failure on customer satisfaction. They investigated whether failures in service operations can change the relative weights that customers assign to key service elements to reach an overall satisfaction assessment. They split their sample into three groups of customers who experience routine service, flight delays of external origin (i.e., weather delays), and flight delays of internal origin (other delays caused by mechanical failures, issues with connecting passengers, and late-arriving aircraft, among many possibilities). Their results show evidence that customer satisfaction is lower for all service failures but suggest that the main components of satisfaction differ between delayed and routine flights only when customers blame the service provider for failure. They also found that internally sourced delays reduce satisfaction more than externally sourced delays and that in the former, employee interactions play a significantly smaller role in customer satisfaction ratings.

## 3. Application

This section presents our empirical study of global passenger satisfaction using survey data. To conduct this study, we develop a methodological proposal to control unobservable factors related to travelers' psychological and situational traits to address the problem of omitted variable bias in respondents' responses. We build an econometric model to examine the formation of passenger satisfaction at São Paulo/Guarulhos International Airport (GRU Airport) in Brazil. Our focus is on the relationship between flight delays and overall passenger satisfaction with the airport and to what extent the possible negative effects of delays can be mitigated by airport administration. In what follows, we present the application, the available data, and the empirical model. In Section 4, we introduce our methodological proposal for modeling and controlling the associated latent factors.

## 3.1. GRU Airport

GRU Airport is located in the metropolitan region of São Paulo city. The metroplex in which it is involved comprises a population of 12 million inhabitants, with a GDP per capita of 15.8 thousand dollars in 2019.[3] It is located 25 km northeast of downtown São Paulo. The main airports in its area of influence are São Paulo/Congonhas (CGH, located 10 km to the south) and Campinas/Viracopos (VCP, located just over 90 km to the north).[4] Figure 1 shows a map of the existing GRU terminals and the airport's geographic location relative to other regional airports and cities.

GRU Airport is one of the main hubs in Latin America and an important gateway for international passengers from all continents. It is the largest airport in Brazil and South America, handling 43 million passengers and 292,000 takeoffs and landings in 2019.[5] It has two runways measuring 3,000 m and 3,700 m in length. It had three passenger terminals. Terminal 1, the smallest, has 11 boarding gates and is generally operated by the Azul Airline for domestic flights. Terminal 2, the oldest and with the largest number of boarding gates (46), operates domestic and international flights to Latin America and has a strong presence of Gol and LATAM airlines. Finally, Terminal 3, the newest and most modern, has 32 boarding gates dedicated to international flight operations, with a wide range of foreign airlines. The airport is coordinated with the IATA level 2 status and handles up to 57 takeoffs and landings per hour. It has 179 parking spaces for aircraft, 54 of which allow boarding with jet bridges.

The airport was privatized on February 6, 2012, with a concession contract valid for twenty years signed in June of that year. On the occasion, the Brazilian government announced that the consortium formed by the companies Invepar and ACSA (Airports Company South Africa) was the winner of the concession auction.

Since privatization, GRU Airport has had its fees fixed under airport regulations carried out by the National Civil Aviation Agency (ANAC). One of the items included in Brazilian airport regulation is monitoring the quality of airport services, which, among other factors, is permanently measured through a Passenger Satisfaction Survey. The survey was conducted through questionnaires with airport passengers, who responded to questions regarding their evaluation of various airport components, such as curbside, check-in processing time, security inspection, quality and variety of retail stores, food and beverage stores, and restaurants, among several other items. Third-party firms subject to specific regulations and permanent audits by ANAC carried out the survey. The results of the evaluations obtained from the survey, along with other collected indicators, compose an airport quality index called the "Q Factor."[6] The Q Factor is a key component of the airport fee regulations in Brazil.

---

[3]Source: IBGE, available at cidades.ibge.gov.br/brasil/sp/sao-paulo/panorama, with own calculations. Dollar values consider an exchange rate of 3.9450 (2019); source: www.ipeadata.gov.br.

[4]Source: Google Maps.

[5]Source: The information in this paragraph was taken from the airport's website, www.gru.com.br, retrieved in November 19, 2022.

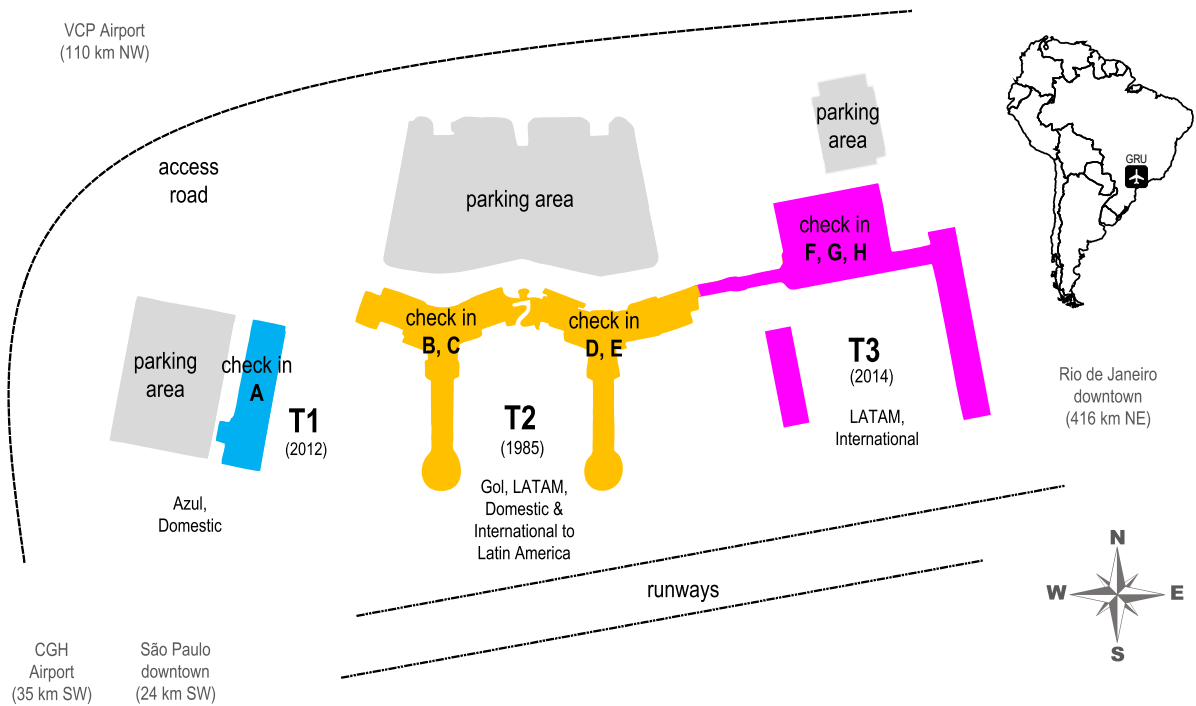

*Source: GRU Airport's website. Authors' elaboration.*

**Figure 1–GRU Airport terminals**

Figure 2 presents images of the boarding areas in the three passenger terminals at GRU airport. In particular, the figure shows boarding gates at times of greater and lesser movement of passengers to provide an idea of the circumstances of the formation of queues for boarding and the possible impacts of delays and flight cancellations on traveler behavior. It is possible to notice differences in the level of service between terminals. Images (i) to (iv) show Terminal 1 (the smallest); images (v) to (viii) show Terminal 2 (the oldest); images (ix) to (xii) show Terminal 3 (the only one exclusively dedicated to international flights). Some images show the presence of retail stores and restaurants. Many of these points of sale are strategically located at the gates close to the queuing areas for boarding. Therefore, we hypothesize that airports benefit from the positioning of these services to increase their global passenger satisfaction level, particularly in flight delays.

---

[6]See updated details in the following regulations: Ordinance N. 3.730/SRA-ANAC, of December 3, 2019, and Ordinance N. 6.059/SRA-ANAC, of September 30, 2021.

Terminal 1

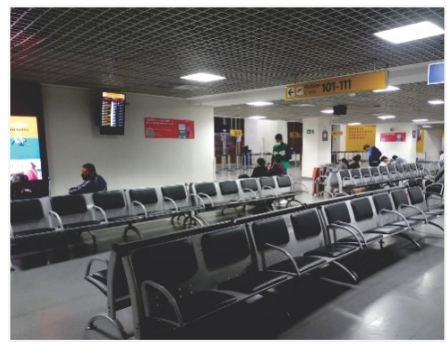
(i)

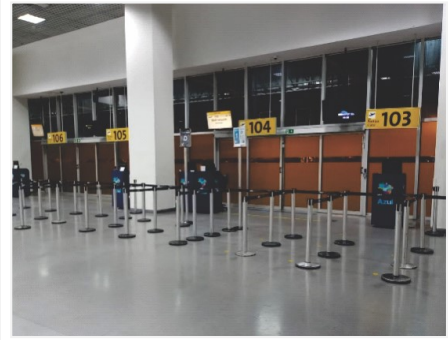
(ii)

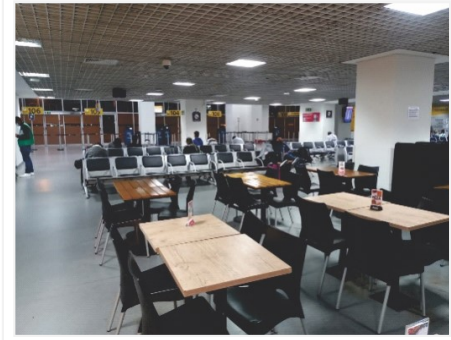
(iii)

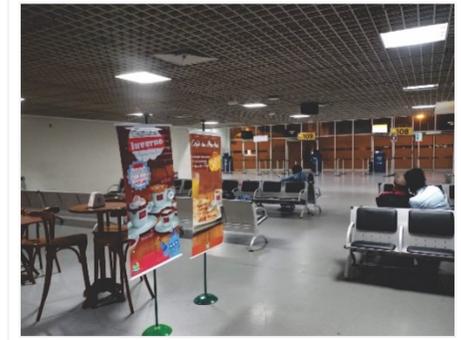
(iv)

Terminal 2

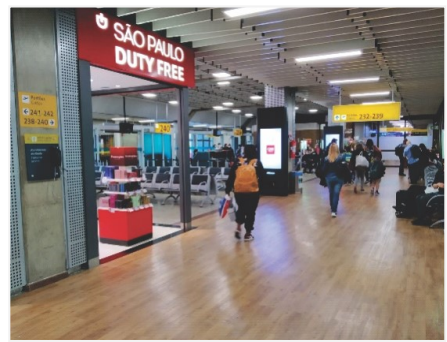
(v)

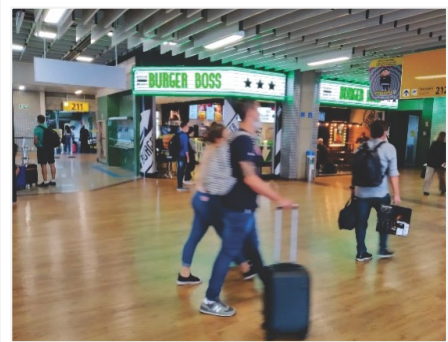
(vi)

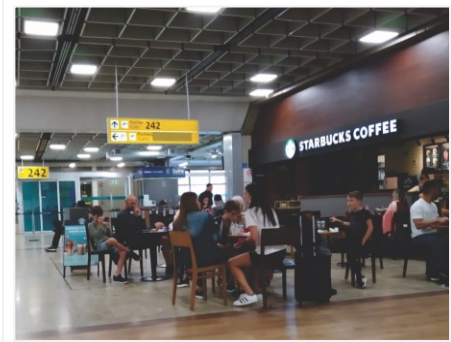
(vii)

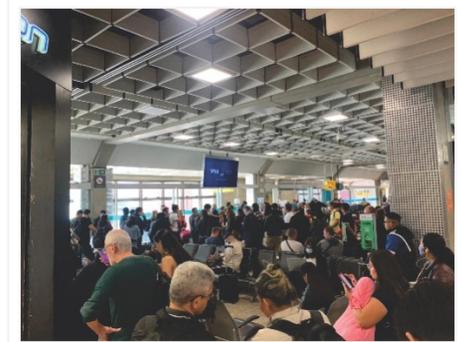
(viii)

Terminal 3

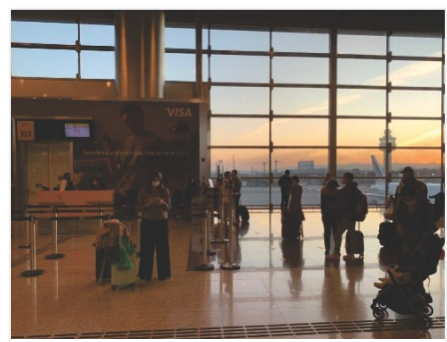
(ix)

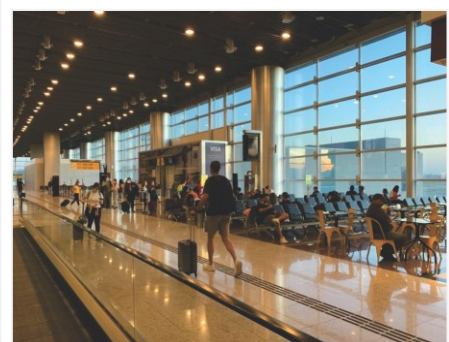
(x)

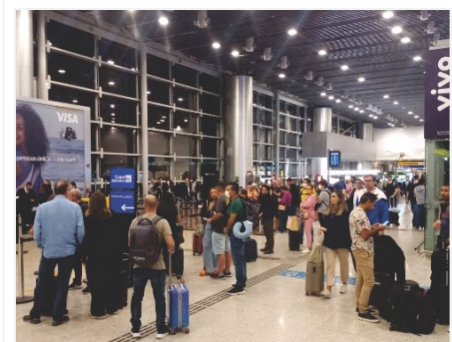
(xi)

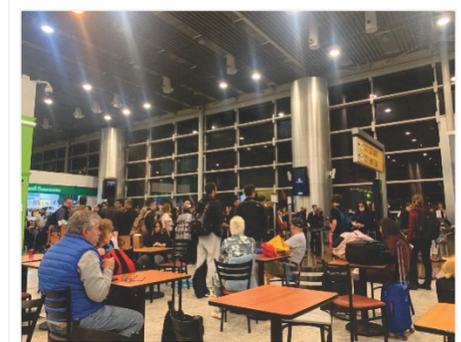
(xii)

Source: the authors.

**Figure 2–Terminal gates of GRU Airport at busy and non-busy hours**

*3.2. Data*

The main data source for this study is the Passenger Satisfaction Survey, under the responsibility of the National Civil Aviation Agency, ANAC – which, from now on, we will denote by the "PS survey." Regulators periodically collect this survey through third-party firms aiming to feed a set of service quality indicators for privatized airports in Brazil as an obligation present in each concession contract. The PS Survey contains several questions related to passenger satisfaction with respect to the components and experiences at the airport and global satisfaction with the service.

In addition to passenger evaluations from the PS Survey, we employed other databases to complement the collected data with operational information on flights, airports, and weather conditions. The different data sources are then interrelated using the flight number indicator and the passenger interview's place, date, and time. We gathered data on departure and arrival times of flights, delays and cancellations, passenger movement, and airport capacity and operations (Source: ANAC). Finally, we collected weather data at airports (Source: Iowa State University Mesonet). In the Appendix, we present a detailed description of the data sources used in our study.

Our database contains 13,071 questionnaires collected from passengers waiting for their flights in the departure areas at Guarulhos airport between February 21, 2018, and July 21, 2021. Of these respondents, 9,133 were domestic passengers, and 3,938 were international passengers. From the original database, interviews conducted over two hours before and after the scheduled flight time were discarded. In addition, interviews with connecting passengers, passengers on canceled flights, or passengers whose flight numbers were missing in ANAC records were discarded. Passengers who would have had subsequent connecting flights were not identified through the questionnaire questions and thus may have been included in the data.

The number of observations according to the year of collection was 4,466 (2018), 5,623 (2019), 1,138 (2020), and 1,844 (2021).[7] Part of our sample relates to the COVID-19 pandemic period, in which flight and passenger movements declined considerably. Additionally, the frequency of flight delays at GRU during this period is less frequent in our sample. In the previous year (2019), the delay rate is around 16.5%. However, starting from March 2020, it decreases to 13.1%, and further drops to 8.4% in 2021. Stores and restaurants at GRU and the other airports in Brazil experienced restricted operations. Some establishments were closed by health authorities due to sanitation concerns. However, there was also a legal dispute regarding the opening of these businesses during this period.

---

[7] Note that the sample sizes for 2020 and 2021 are considerably smaller compared to 2018 and 2019. The samples from 2020 and 2021 account for only 8.7% and 14.1% of the total observations, respectively. This disparity can be attributed to the interruption of the research in March 2020 due to the pandemic outbreak, with data collection resuming in January 2021. Furthermore, the current sample for 2021 only includes data up until July.

### 3.3. Flight delays and passenger satisfaction statistics

In recent years, Brazilian air transport statistics indicate mean delay rates between 6 and 8 percent of the total scheduled flights between 2018 and 2020.[8] However, official data reported by ANAC assumes that a flight is considered delayed if its arrival takes place more than 30 min after the scheduled time. If we consider delays equal to or greater than 15 minutes, such as the US Department of Transportation, these percentages rise to the range between 11 and 16 percent.[9] For comparison, the average delay in air transport is between 10 and 20 percent over the same period. GRU Airport showed percentages of delays systematically above the national average between 2018 and 2020, with values ranging between 12 and 19 percent, but with the lowest values observed during the COVID-19 pandemic.[9]

Figure 3 illustrates the relationship between flight delays and the perceived quality of the service provided by the GRU Airport. We built many graphs displayed in the figure from ANAC's Passenger Satisfaction Survey data collected at the airport between 2018 and 2021. The figure shows the behavior of the overall passenger satisfaction scores within the 10-point Likert scale, which is the scale adopted in the survey questionnaires.[10] We show the results of the sample means of these evaluations, according to a varied portfolio of standpoints, referring to the characteristics of the interviewed passengers (generation, schooling, trip frequency, satisfaction with restaurants, and satisfaction with shops) and the characteristics of their flights (departing airport terminal, flight destination, weather conditions, time until scheduled departure, and actual flight delay duration).

As shown in Figure 3, the sample means of the passengers' global satisfaction scores were generally between 7 and 8. However, note that in all cases, these scores decrease in situations of flight delays, as we compare the blue bars with the orange bars. For example, among passengers belonging to Generation X, the mean score is 7.91 when passenger flights depart on time. The mean score drops to 7.65 (down 3.3%) in cases of delayed departure. We observed a similar effect in the evaluations of passengers of other generations, schooling, and trip frequency. Another example concerns the weather conditions. Both in situations of favorable weather conditions ("Ceiling and Visibility OK") and in situations that the weather is not so favorable ("Ceiling and Visibility Not OK"), the materialization of a flight delay leads to a drop in the mean rating of passengers: from 8.09 to 7.97 in the first case (-1.5%), and from 8.10 to 7.92 in the second case (-2.2%). Several other analyses are possible from the graphs in Figure 3 when contrasting routine and delayed flights.

---

[8]Source: ANAC's Air Transport Yearbook (2021), for the period between 2018 and 2021. See www.nectar.ita.br/avstats/anac_yearbook.html for a description and a link to the original website in Portuguese.

[9]Source: ANAC's Active Scheduled Flight Historical Data Series, for the period between 2018 and 2020, with own calculations. See www.nectar.ita.br/avstats/anac_vra.html.

[10]These data constitute the sample used in the present study. See details in 3.2.

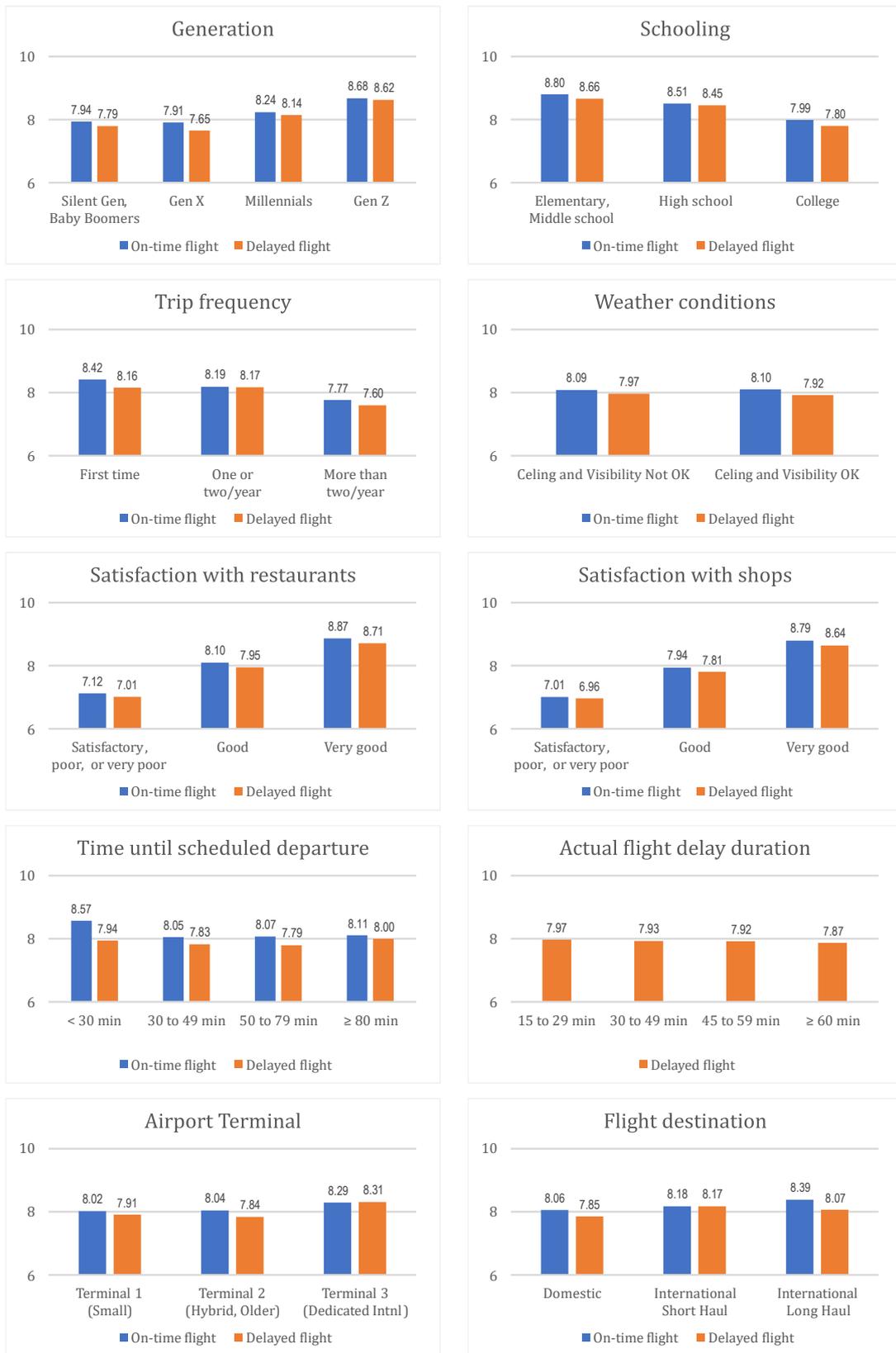

*Source: Passenger Satisfaction Survey ("PS Survey") of GRU Airport, from February 2018 to July 2021, with own calculations. Passenger satisfaction measured on a 10-point Likert scale. See 3.2 for a detailed discussion of the sources and data.*

**Figure 3–Mean airport passenger global satisfaction rates at GRU Airport–on-time vs. delayed flights**

*3.4. Empirical model*

One of the questions asked in the PS Survey interviews with GRU Airport passengers and included in our data sample is "*In general, I am satisfied with this airport (rate from 1 to 10)*" (source: PS Survey). The question accepts responses on a 10-point Likert scale, with anchors of "strongly disagree" and "strongly agree."[11] Based on the responses to this question, we created an airport passenger global satisfaction variable (APTSAT).

To investigate the relationship between flight delays and APTSAT, we propose the following econometric framework estimated using an ordered probit model:

$$\begin{aligned}
\Pr(\text{APTSAT} = i) = \Pr(\kappa_{i-1} \\
< \beta_1 \text{GENSILEN} + \beta_2 \text{GENBOOM} + \beta_3 \text{GENMILLEN} + \beta_4 \text{GENZ} \\
+ \beta_5 \text{SCHLELEM} + \beta_6 \text{SCHLMIDD} + \beta_7 \text{SCHLHIGH} + \beta_8 \text{FIRSTTFLIER} \\
+ \beta_9 \text{FREQFLIER} + \beta_{10} \text{LSRFLIER} + \beta_{11} \text{INTNLDEST} + \beta_{12} \text{REDEYE} \\
+ \beta_{13} \text{SMALLTERM} + \beta_{14} \text{INTNLTERM} + \beta_{15} \text{TERMDEN} + \beta_{16} \text{JETBRIDGE} \\
+ \beta_{17} \text{SHOPS} + \beta_{18} \text{FOOD} + \beta_{19} \text{EXPENSIVE} + \beta_{20} \text{WIFI} + \rho\,\text{DEL} + u \leq \kappa_i)
\end{aligned} \quad (1)$$

where APTSAT is the global passenger satisfaction rate of the airport, $i$ is each possible outcome for APTSAT, and $\kappa_i$, $i = \{1, 2, \ldots, I-1\}$, represents the $I-1$ cutpoints to be estimated, along with the model parameters ($\beta$s). In our case, as the survey employed a 10-point scale, $I = 10$. For estimation purposes, the ordered probit procedure requires setting extreme cut-off points equal to $\{\kappa_0, \kappa_I\} = \{-\infty, +\infty\}$. Note that $\rho$ is the parameter of interest as it models the possible negative association between flight delays (DEL) and passenger satisfaction (APTSAT).

The covariates used in the model in Equation (1) can be classified into the following categories:

- PAX profile: generation: GENSILEN (dummy variable indicating that the passenger belongs to the Silent Generation), GENBOOM (Baby Boomer Generation), GENMILEN (Millennial Generation), and GENZ (Z Generation); the reference cases for these dummies are passengers classified as belonging to Generation X (GENX).

- PAX profile: schooling: SCHLELEM (dummy variable indicating that the passenger has completed up to the fourth year of elementary school), SCHLMIDD (complete middle school), SCHLHIGH (complete high school); the reference cases for these dummies are passengers classified as having higher education, complete or incomplete (SCHLCOLL).

---

[11] This information served as the basis for the construction of the graphs in Figure 3.

- <u>PAX profile: travel experience and purpose</u>: FIRSTTFLIER (dummy indicating that the passenger is traveling by plane for the first time), FREQFLIER (passengers who fly frequently, more than three times a year), LSRFLIER (passengers traveling for leisure); the reference cases for the dummies are passengers who travel once or twice by plane in the year prior to the survey (EXPERCDFLIER). For LSRFLIER, the reference case is passengers traveling for business purposes (BSNFLIER) and other travel motives.

- <u>Flight characteristics</u>: INTNLDEST (dummy variable to indicate that the passenger is traveling to an international destination) and REDEYE (dummy variable to indicate an overnight flight);

- <u>Airport experience: boarding</u>: SMALLTERM (dummy variable to indicate that the passenger is boarding at Terminal 1), INTNLTERM (dummy variable to indicate that the passenger is boarding at Terminal 3); the reference case for these dummies is Terminal 2, the oldest and hybrid domestic/international; TERMDEN (passenger density at the terminal where the passenger answers the questionnaire), JETBRIDGE (dummy variable to indicate whether the passenger's flight uses a boarding bridge)

- <u>Airport experience: products/services</u>: SHOPS (passenger satisfaction scores with the quality and variety of stores at the airport), FOOD (passenger satisfaction scores with the quality and variety of foods), EXPENSIVE (passenger satisfaction scores with the price of shops and food at the airport), and WIFI (satisfaction with the WI-FI service).

- <u>Flight delays</u>: DEL (dummy variable indicating whether the flight had its takeoff delayed by more than 15 minutes).

Finally, $u$ is the composite error term, which is discussed in detail in the next subsection. More details about the variables and their construction are provided in the Appendix. Table 1 presents descriptive statistics of the variables used in this study.[12]

## 4. Methodology

In this section, we present our methodology. The basic idea of the modeling is to perform econometric control of passengers' psychosituational traits when utilizing the data from the PS survey. Our methodology is motivated by the meta-theoretical framework of Mowen (2000), which consists of relationships between the variables of behavior, situational context, and the personality of individuals. In the author's approach, the use of psychological trait variables played a crucial role in the empirical study of these relationships. In this framework, a "trait" is defined as "*any intra-psychic construct that can be measured validly and reliably, and that predicts individual differences in feelings, thoughts, and behaviors.*"[13]

---

[12] Note that we have included various other variables in Table 1 that are not part of Equation (1) but will be utilized for conducting robustness checks of the model. We will present and discuss these additional variables in more detail in subsequent sections and in the Appendix.

However, our approach differs from that of Mowen (2000) because of the indirect treatment of trait modeling. We start from the assumption of the existence of a commonly observed limitation in the satisfaction survey data, dictated by the non-observability of psychological nature variables, which prevents us from having explicit trait variables. However, when we hypothesize the process that generates these data, we recognize the possible validity of internal interactions between these variables, as modeled by the author. Additionally, we assumed the validity of the possible associations of omitted psychological factors with some of the variables included in our empirical model. Thus, in the absence of a more direct approach to intra-psychic constructs and their effects on human behavior, we suggest, as an alternative, the use of an indirect specification approach in our passenger satisfaction model through proxy and control variables for latent traits.

### *4.1 Psychosituational latent traits*

Our empirical framework dictated by Equation (1) is a baseline model that contains a set of possibly pertinent variables but still presents some key econometric challenges. In particular, we recognize its limitations arising from the existence of unobservable factors for the econometrician that may influence passenger satisfaction. Consistent with Mowen (2000), we assume that most of these unobservable factors are related to the personality of the respondents (i.e., psychological characteristics of more enduring patterns), their emotional context (i.e., more situational psychological issues dictated by the environment), and the interaction between these two factors. Additionally, it is likely that these unobservables also suffer influence from the possible airline interventions in the period, such as calls and public announcements, especially as the scheduled boarding and flight time approaches. All these elements constitute what we denote here as *latent psychosituational traits* that can influence the responses during the face-to-face survey at the airport. The existence of these traits causes a potential bias in omitted variables in our modeling.

---

[13]Mowen (2000), p. 2.

**Table 1–Descriptive statistics of model variables**

| Variable | Brief description | Metric | Mean | Std. Dev. | Min | Max |
|---|---|---|---|---|---|---|
| AIRCSIZE | number of aircraft seats | count | 2.04 | 0.67 | 0.50 | 5.16 |
| APTSAT | satisfaction with airport | 10-points scale | 8.07 | 1.72 | 1.00 | 10.00 |
| BOARD (CALL) | interview - boarding call was imminent | dummy | 0.74 | 0.44 | 0.00 | 1.00 |
| BOARD (NOT) | interview - boarding not commenced/not imminent | dummy | 0.02 | 0.13 | 0.00 | 1.00 |
| BOARD (NOW) | interview - boarding has likely commenced | dummy | 0.24 | 0.43 | 0.00 | 1.00 |
| BUSYDAY | number of daily passengers at terminal | count/10,000 | 2.92 | 1.36 | 0.18 | 4.94 |
| BUSYHOUR | number of hourly passengers at terminal | count/1,000 | 1.61 | 1.05 | 0.00 | 4.98 |
| CARGO | cargo load | tons/10 | 0.27 | 0.53 | 0.00 | 5.50 |
| CASCAD (ARR) | cascading delay (arrival) | proportion | 0.09 | 0.09 | 0.00 | 0.75 |
| CASCAD (DEP) | cascading delay (departure) | proportion | 0.15 | 0.15 | 0.00 | 0.95 |
| DEL | flight delay of more than 15 minutes | dummy | 0.17 | 0.37 | 0.00 | 1.00 |
| DELDUR | flight delay duration of more than 15 minutes | hours | 0.13 | 0.42 | 0.00 | 5.97 |
| DISSAT (AIRLINE) | dissatisfaction with airline staff (check in) | ratio | 1.78 | 2.63 | 0.00 | 22.00 |
| DISSAT (CHECKIN) | dissatisfaction with check-in waiting time | ratio | 1.50 | 1.66 | 0.00 | 13.13 |
| DISSAT (CURBSID) | dissatisfaction with the terminal curbside | ratio | 1.75 | 1.80 | 0.00 | 12.50 |
| DISSAT (FLTINFO) | dissatisfaction with flight information | ratio | 1.06 | 1.17 | 0.00 | 7.00 |
| DISSAT (SECINSP) | dissatisfaction with security inspection time | ratio | 1.07 | 1.16 | 0.00 | 11.00 |
| DISSAT (WALKDST) | dissatisfaction with terminal walking distance | ratio | 1.04 | 0.70 | 0.00 | 4.54 |
| DISSAT (WAYFIND) | dissatisfaction with terminal wayfinding | ratio | 1.02 | 1.08 | 0.00 | 6.48 |
| DISTANCE | flight distance | miles/1000 | 1.64 | 1.82 | 0.18 | 7.59 |
| EXPENSIVE | shops/food high price perception | [0,1] index | 0.52 | 0.26 | 0.00 | 1.00 |
| EXPERCDFLIER | experienced flier | dummy | 0.30 | 0.46 | 0.00 | 1.00 |
| FIRSTTFLIER | first-time flier | dummy | 0.31 | 0.46 | 0.00 | 1.00 |
| FOOD | satisfaction with quality/variety of food | [0,1] index | 0.70 | 0.28 | 0.00 | 1.00 |
| FOOD (4/5 RATING) | quality/variety of food rating of 4 or 5 | dummy | 0.53 | 0.50 | 0.00 | 1.00 |
| FREQFLIER | frequent flier | dummy | 0.39 | 0.49 | 0.00 | 1.00 |
| GENBOOM | Baby-Boomer Generation | dummy | 0.17 | 0.37 | 0.00 | 1.00 |
| GENMILLEN | Millennial Generation | dummy | 0.40 | 0.49 | 0.00 | 1.00 |
| GENSILEN | Silent Generation | dummy | 0.04 | 0.19 | 0.00 | 1.00 |
| GENX | Generation X | dummy | 0.32 | 0.47 | 0.00 | 1.00 |
| GENZ | Generation Z | dummy | 0.07 | 0.25 | 0.00 | 1.00 |
| INTNLDEST | international destination | dummy | 0.31 | 0.46 | 0.00 | 1.00 |
| INTNLTERM | international terminal (Terminal 3) | dummy | 0.22 | 0.42 | 0.00 | 1.00 |
| JETBRIDGE | jet bridge boarding | dummy | 0.77 | 0.42 | 0.00 | 1.00 |
| LOADFAC | load factor of flight | proportion | 0.82 | 0.17 | 0.04 | 1.00 |
| LSRFLIER | travel purpose - leisure | dummy | 0.62 | 0.49 | 0.00 | 1.00 |
| PANDEMIC (EARLY) | period from Mar to Dec 2020 | dummy | 0.03 | 0.17 | 0.00 | 1.00 |
| PANDEMIC (LATER) | period from Jan to Dec 2021 | dummy | 0.14 | 0.35 | 0.00 | 1.00 |
| PANDEMIC (PRE) | period from Jan to Dec 2019 | dummy | 0.43 | 0.50 | 0.00 | 1.00 |
| PRCONNECT | proportion of connecting passengers | proportion | 0.05 | 0.12 | 0.00 | 1.00 |
| REDEYE | red-eye flight | dummy | 0.12 | 0.32 | 0.00 | 1.00 |
| RUNWAYCONG | runway congestion | proportion | 0.61 | 0.19 | 0.02 | 1.16 |
| RUNWAYDIS | runway disruption | proportion | 0.13 | 0.12 | 0.00 | 0.86 |
| SCHLCOLL | schooling - college | dummy | 0.81 | 0.39 | 0.00 | 1.00 |
| SCHLELEM | schooling - elementary | dummy | 0.02 | 0.13 | 0.00 | 1.00 |
| SCHLHIGH | schooling - high school | dummy | 0.14 | 0.35 | 0.00 | 1.00 |
| SCHLMIDD | schooling - middle school | dummy | 0.03 | 0.16 | 0.00 | 1.00 |
| SHOPS | satisfaction with quality/variety of shops | [0,1] index | 0.70 | 0.30 | 0.00 | 1.00 |
| SHOPS (4/5 RATING) | quality/variety of shops rating of 4 or 5 | dummy | 0.58 | 0.49 | 0.00 | 1.00 |
| SMALLTERM | small terminal (Terminal 1) | dummy | 0.08 | 0.27 | 0.00 | 1.00 |
| TERMDEN | terminal density | pax/area × 10 | 0.23 | 0.15 | 0.00 | 1.00 |
| TERMDIS | proportion of disrupted flights of a terminal | proportion | 0.14 | 0.20 | 0.00 | 1.00 |
| WEATHER (DST) | weather-related issues (destination) | dummy | 0.09 | 0.28 | 0.00 | 1.00 |
| WEATHER (ORG) | weather-related issues (origin) | dummy | 0.03 | 0.16 | 0.00 | 1.00 |
| WIFI | satisfaction with airport Wi-Fi service | [0,1] index | 0.45 | 0.38 | 0.00 | 1.00 |
| WIFI (4/5 RATING) | quality of Wi-Fi service rating of 4 or 5 | dummy | 0.36 | 0.48 | 0.00 | 1.00 |

*Note: See the Appendix for a detailed description of the variables and their sources.*

The problem of latent (omitted) variables can raise issues of endogeneity and biased estimation of the marginal effect of DEL on APTSAT in Equation (1). First, fixed psychological traits (e.g., optimism vs. pessimism) can influence how APTSAT is generated. For example, it is intuitive to think that pessimistic (optimistic) individuals may be more (less) dissatisfied with the airport in situations of flight delays due to mood issues and anxiety in anticipating negative future experiences.[14] Consistent with Mowen (2000), these fixed trait factors can be nested within a broader classification of psychosituational factors that can be fixed or variable. Among the variables for psychosituational factors, we have, for example, passengers emotionally affected by urban traffic problems caused during their journey to the airport or those who arrived late at the terminal and, because of the rush, did not have a smooth way to their boarding gate. These passengers may be more likely to take out such an emotional state in their evaluation of the airport. Another example would be passengers who had previous dissatisfaction in their relationship with the airline, such as during online customer service or check-in counters, who may also be more likely to complain about airport service. In any of these examples, the idiosyncratic elements of the passenger's personality may interact with the situational context, affecting their behavior (à la Mowen, 2000) and, ultimately, influencing their response to the survey questionnaires.

To analyze the causal relationship between flight delays and passenger satisfaction, with an unbiased estimation of the ceteris paribus marginal effect of DEL in Equation (1), it is fundamental that the possible unobservable factors of APTSAT that are correlated with this variable are at least controlled to make the DEL mean independent of the error term $u$. Therefore, we need a specific modeling strategy to address this potential source of endogeneity that contaminates the estimates.

Given that the specification of the model aims to measure the influence of latent psychological characteristics, we explore the associations that emerge in the problem through the path diagram depicted in Figure 4. This approach typically provides a framework for analyzing the complex relationships among variables in a comprehensive manner. In our current study, we utilize a path diagram as a visualization strategy to gain insights for our modeling process.[15]

The first relationship we focus on in Figure 4 is one of the most straightforward: the measured variable of overall satisfaction rating of an airport attributed by a passenger ("Global Rating") is directly influenced by the latent variable of actual satisfaction with the airport ("Airport Satisfaction"). This relationship is represented by the directional arrow connecting the variables in the bottom left corner of the diagram. Additionally, "Global Rating" is also directly influenced by another latent variable, "Blame Attribution." This means that the passenger may attribute blame for a flight delay to the airport, consequently affecting their declared airport satisfaction rating.

---

[14] As in studies of humor by Wright & Bower (1992), and MacLeod & Byrne (1996).

[15] However, we recommend that future studies consider a more in-depth structural equation modeling (SEM) approach as a possible strategy to further enhance the understanding of the latent constructs and their relationships. SEM offers a robust framework for incorporating the measurement model and conducting path analysis, allowing for a more comprehensive examination of the variables involved.

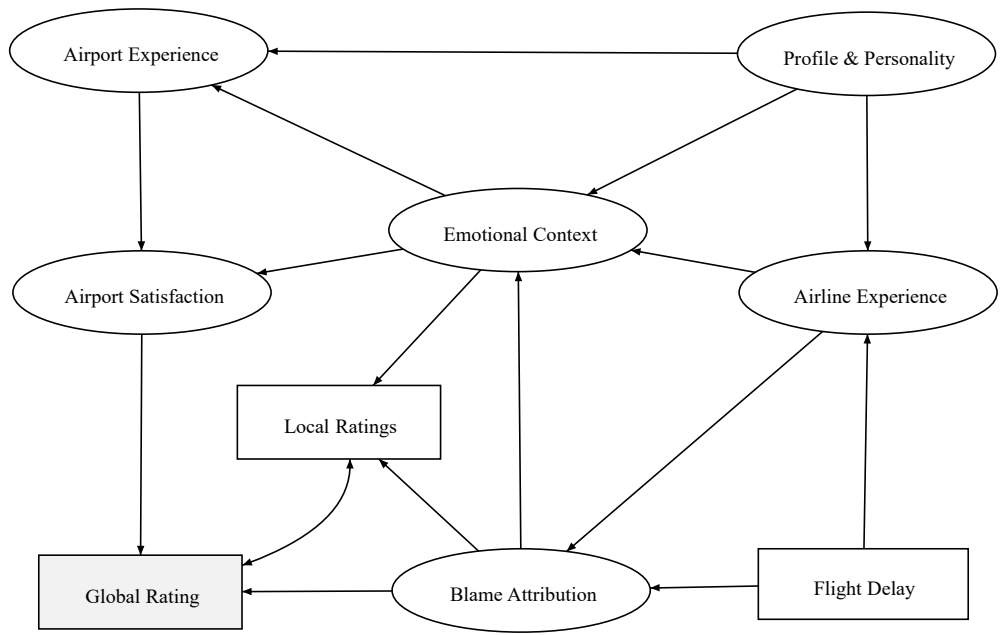

**Figure 4–Path diagram illustrating model relationships**

Note the curved arrow connecting "Global Rating" to "Local Ratings." This relationship builds upon the recommendations of van Praag and Ferrer-i-Carbonell (2008) for leveraging the common personality traits that influence responses to questions regarding the satisfaction with global and local aspects of an individual's experience. In the authors' approach, the corresponding terms of "global ratings" and "local ratings" are the concepts of "general satisfaction" and "domain satisfactions," respectively.[16] These concepts are integrated into an equation model through an "aggregating approach".[17] We represent this relationship with a curved line with double arrows at each end, indicating a covariance between the two variables. This signifies a non-directional path, indicating that we know the two variables are associated.

At the center of the diagram in Figure 4, we have the latent variable "Emotional Context," which is influenced by the "Profile & Personality" of the passenger. Additionally, we model "Emotional Context" as being influenced by the "Airline Experience," encompassing any previous experiences the passenger had with the airline, including aspects such as the purchasing process, check-in procedures, and other relevant interactions. It is worth noting that the "Flight Delay" may also influence the "Airline Experience" of the passenger, and ultimately, their emotional context. Furthermore, "Blame Attribution" acts as a determinant of "Emotional Context." In turn, "Emotional Context" plays a significant role in driving both "Airport Experience" and "Airport Satisfaction."

---

[16] See, for example, van Praag and Ferrer-i-Carbonell (2008, Chapters 1 and 4).

[17] van Praag and Ferrer-i-Carbonell (2008, p. 86).

We acknowledge that the path diagram presented in Figure 4 provides a simplified representation of the complexity surrounding the formation of passengers' airport satisfaction. However, it serves to illustrate two key issues. Firstly, it suggests that although "Flight Delays" does not have a direct relationship with latent psychosituational traits, it may exert an indirect influence on "Airport Satisfaction" through "Blame Attribution" and "Emotional Context." In our framework, "Emotional Context" and "Profile & Personality" are crucial elements of the latent traits that need to be controlled for in order to avoid omitted variable bias. Secondly, the diagram allows for the visualization of the relationship between "Emotional Context," "Local Ratings," and "Global Rating". Specifically, in the diagram, we observe that "Emotional Context" acts as a determinant of the "Local Ratings" declared by the passenger. This relationship allows us to incorporate econometric controls for the psychosituational latent traits of passengers within the APTSAT equation. We achieve this by leveraging their reported satisfaction with the various components/stages they have previously experienced at the airport, as we will discuss in further detail below.

The core of our methodological proposal is related to the following empirical strategy based on van Praag and Ferrer-i-Carbonell (2008), with adaptations and extensions. To address the complexity of the phenomenon generated by latent psychosituational traits in Equation (1), we treat $u$ as a composite error. In this sense, the first task is to recognize the variability of $u$ in several dimensions relevant to the passenger satisfaction formation phenomenon, as expressed in Equation (2).

$$u = u_{iadh}, \qquad (2)$$

where $i$, $a$, $d$, and $h$ refer to the individual, airport, flight date, and interview time, respectively. By defining $u$ in this manner, we recognize the existence of unobserved idiosyncrasies associated with these dimensions. Our proposed specification of $u_{iadh}$ is:

$$u_{iadh} = \Omega_1(\xi_i, \xi_{idh}) + \Omega_2(\xi_j, \xi_{jdh}) + \Omega_3(\xi_a, \xi_{adh}) + \Omega_4(\xi_d, \xi_h) + \omega_{iadh}, \qquad (3)$$

where the $\Omega$'s are unknown functions, and the $\xi$'s are unobserved factors that may be specific to each dimension but may also vary over two or more dimensions. For example, $\xi_{idh}$ indicates unobserved factors that are individual-date-hour specific. $\omega_{iadh}$ is a conventional, normally distributed error term. Note that it $\Omega_2$ contains unobserved factors related to the airline (denoted by index $j$), meaning that passenger satisfaction may be affected by idiosyncrasies specific to the carrier ($\xi_j$), for example, its mean perceived reputation. We also assume that there may be carrier factors related to passenger perceptions of its efficacy in managing flight operations at the time of the survey ($\xi_{jdh}$). Through these abstract components of (3), we first approximated respondents' latent psychosituational traits.

However, to enable the estimation of Equation (1) using the proposal in (3), it is essential to specify the function Ω in a way that allows its econometric handling. We then propose the following simplification, obtained using proxies and control variables. First, we propose the following factors embedded in Ω, according to (4):

$$u_{iadh} = \Omega_1(\text{DISSAT}_{iadh}, \text{DEST}^{(r)}) + \Omega_2(\text{AIRL}^{(j)}, \text{DISSAT}_{ijdh}) \quad (4)$$
$$+ \Omega_3(\text{DISSAT}_{iadh}, \text{TERMDIS}_{a_m dh}) + \Omega_4(\text{DATE}^{(d)}, \text{TIMEOFLT}_h)$$
$$+ \omega_{iadh},$$

where:

- $\text{DISSAT}_{iadh}$ it is a set of controls for passenger dissatisfaction with the airport from arrival at the airport to the moment of the interview.

- $\text{DISSAT}_{ijdh}$ it is a control of the passenger's dissatisfaction with the airline;

- $\text{TERMDIS}_{a_m dh}$ serves as an indicator of passenger congestion on the airport concourse within terminal $a_m$, where $m$ is the terminal number at GRU Airport, $m = \{1,2,3\}$, particularly among travelers on delayed flights awaiting boarding announcements. It denotes the proportion of passengers on delayed flights waiting for boarding at the time of passenger interviews.

- $\text{DEST}^{(r)}$, $\text{AIRL}^{(j)}$, and $\text{DATE}^{(d)}$ are sets of control variables (dummies) designed to account for the idiosyncrasies of the destination, airline, and date of the passenger's trip within the survey.

- $\text{TIMEOFLT}_h$ is a set of control variables (dummies) of the time intervals from the moment of flight.

Another important aspect to consider is the psychosituational latent traits of passengers during the Covid-19 pandemic. With the pandemic in full swing, many individuals chose to avoid non-essential travel, leading to a higher proportion of air travelers with urgent and critical reasons for their journeys. As a result, flight delays during this time may have considerably heightened the dissatisfaction experienced by these individuals. It is crucial to acknowledge that the traveling experience and conditions in airports underwent significant variations during this period, which could have exerted an influence on the psychological aspects of air travel. To account for these factors, we have introduced dummy variables for the pandemic in our analysis, allowing us to control for these unique circumstances and explore their impact on passenger satisfaction. By incorporating these controls, we are able to account for the overall average impact of the flight date dummies, which retain their date-specific interpretation. The pandemic dummies we have included are PANDEMIC (PRE), PANDEMIC (EARLY), and PANDEMIC (LATER), representing the year prior to the pandemic onset (2019), the early months of the pandemic in 2020 until the end of that year, and the year 2021, respectively.[18]

The final specification of $u_{iadh}$ is therefore equal to:

$$u_{iadh} = \sum_k \delta_{1k} \text{DISSAT}_{iadh}^{(k)} + \delta_2 \text{DISSAT}_{ijdh} + \delta_3 \text{TERMDIS}_{adh} \quad (5)$$
$$+ \sum_l \delta_{4l} \text{TIMETOFLT}_h^{(l)} + \text{DEST}_r + \text{AIRL}_j + \text{DATE}_d$$
$$+ \text{PANDEMIC}(PRE)_d + \text{PANDEMIC}(EARLY)_d$$
$$+ \text{PANDEMIC}(LATER)_d + \omega_{iadh}.$$

We used the following covariates in the specification of Equation (5): DISSAT (CURBSID), DISSAT (CHECKIN), DISSAT (WAYFIND), DISSAT (WALKDST), DISSAT (FLTINFO), DISSAT (SECINSP), DISSAT (AIRLINE), TERMDIS, in addition to of dummies referring to TIMETOFLT, DEST, AIRL, and DATE. See Table 1 for a brief description of the variables and Appendix for details of their construction.

We developed DISSAT variables based on the recommendations of van Praag and Ferrer-i-Carbonell (2008). The authors' framework takes advantage of a characteristic commonly observed in satisfaction surveys: the fact that there are different questions about different domains of satisfaction posed to the same respondent[19]. These questions aim to map passengers' local satisfaction compared to their global satisfaction evaluations. In these situations, the psychological traits of respondents can affect their responses to each of the different domains under evaluation similarly. Thus, the modeling explicitly assumes a *common personality effect* that emerges in response to global and local satisfaction.[20]

Therefore, our DISSAT covariates originate from the survey, namely questions concerning satisfaction with the components/stages experienced at the airport.[21] For example, DISSAT (CURBSID) is a variable that comes from the following survey question: '*How do you rate the ease*

---

[18] We thank two anonymous reviewers for suggesting the inclusion of these variables.

[19] In this sense, each component of the whole being evaluated can be considered a "domain."

[20] See Cárdenas et al. (2009) for an exposition and application.

[21] Note that, unlike van Praag and Ferrer-i-Carbonell (2008), who use satisfaction metrics obtained from passengers, here we prefer to use an inverted scale, that is, "dissatisfaction." We believe that, when it comes to providing services, talking about passenger dissatisfaction – rather than satisfaction – is a more intuitive interpretation. Changing the scale does not affect the empirical results of interest, however. Another difference between our approach and that of the authors is that we did not use principal component analysis to isolate the common psychological effects of the covariates (because they are difficult to interpret), and we opted for the use of the ordered probit model, which is the most popular model among survey researchers.

*of entering or leaving the vehicle on the access road next to the terminal entrance (curbside)?"*[22] This variable aims to control passenger dissatisfaction with the curb at the airport. However, it also possibly serves as a proxy for their dissatisfaction with urban traffic congestion during and until arrival at the airport relative to the dissatisfaction of other passengers. We believe that other aspects of passengers' personalities and emotional context may also be correlated with the response to this question.[23]

The other DISSAT variables have similar modeling goals, aiming to control the relative dissatisfaction of the passenger with other stages of the service provision: check-in (CHECKIN), wayfinding (WAYFIND), walk distance (WALKDST), flight information (FLTINFO), security inspection (SECINSP), and airline services (AIRLINE). In all cases, we used van Praag and Ferrer-i-Carbonell's (2008) assumption of the existence of common factors in satisfaction responses. These common factors would generate both APTSAT and DISSAT, which would otherwise be included in the residual term. In this case, the latent psychological traits would make passengers more (less) likely to express greater (less) global satisfaction with the airport than the mean passenger. Thus, the inclusion of these regressors may capture these otherwise omitted factors.

The van Praag and Ferrer-i-Carbonell (2008) approach allows controlling for the common effects of dissatisfaction present in the passengers' responses, being able to purge the respondents' latent *fixed personality traits from the error term.* However, Mowen's (2000) meta-theoretical framework predicts the interaction between psychological and situational factors in the generation of human behavior. To make our approach consistent with Mowen (2000), we extended the van Praag and Ferrer-i-Carbonell (2008) approach, assuming the possibility of interactions between personality and the environment before and during the interview, to control for *variable* latent psychosituational factors that would otherwise be present in the residuals. In our proposal, these factors are considered in the DISSAT variables, which are computed considering the passenger rating relative to the mean ratings of the other passengers interviewed at the time of the interview. Thus, DISSAT has not only variability over the individual-specific dimension but also over the respondents' collectivity and the conjunctural state of affairs at the airport in that given period.

---

[22] Source: PSP/ANAC, with own calculations.

[23] van Praag and Ferrer-i-Carbonell (2008, p. 89) discuss a potential endogeneity bias in estimating the relationship between general satisfaction and domain satisfaction. In our case, to partially mitigate this issue, we only consider questions related to components previously utilized by travelers during their journey from the airport curbside to the boarding gate. However, relying on past experiences preceding the interview as predetermined and less correlated with the unobserved components of the passenger's global rating is a limiting assumption, which we recommend extending in future studies. In any case, we believe that estimation issues related to the DISSAT variables do not diminish their role as nuisance controls in the satisfaction model, and their presence as regressors helps improve the estimation of the effects of the other covariates in the equation.

Other psychosituational variables that extend van Praag and Ferrer-i-Carbonell's (2008) model are TERMDIS and TIMETOFLT. TERMDIS is concerned with controlling the psychological state of passengers in situations of higher incidence of flight delays at the airport. Unlike TERMDEN, TERMDIS aims to capture an additional element of passenger stress, possibly caused by crowding at the airport terminal emerging from a greater number of flights being simultaneously delayed. In this sense, TERMDIS aims to control the emotional context of the passenger in relation to passenger agglomeration at the terminal, as well as the state of tension, irritation, anxiety, etc., of other passengers at the time of the interview. With TERMDIS, our model accommodates the possibility of having some kind of negative spillover from the dissatisfaction of other passengers on delayed flights at the time of the interview on the responses the interviewed passenger gave.[24]

Finally, DEST, AIRL, and DATE are a set of dummy variables inserted into the model to control several other unobservable factors correlated with the trip's destination, the airline's, and the flight's date, respectively.

An important aspect of our methodology is the flexible model selection procedure. Note that to estimate the ordered probit model of satisfaction considering the latent psychosituational traits of the passenger, Equation (5) must be plugged into Equation (1). As Equation (5) deals with latent factors unobserved by the econometrician, we must assume that some (or many) of the proposed controls may not be relevant in explaining passenger satisfaction. Given that the number of proposed controls is relatively large (393), we employed a procedure to select the covariates to be incorporated into Equation (1). For this purpose, we utilize the high-dimensional sparse (HDS) regression models of Belloni et al. (2012), Belloni, Chernozhukov, and Hansen (2014a, b), and Chernozhukov, Hansen, and Spindler (2015). These models build upon Tibshirani's (1996) Least Absolute Shrinkage and Selection Operator (LASSO).

High-dimensional sparse regression is a statistical challenge that arises when the number of predictors in a model is substantial. This situation is not uncommon across various fields, particularly when the true model is considered unknown. As datasets continue to grow, the use of many controls is becoming increasingly common. Regularized regression methods, such as LASSO, are designed to handle models with numerous controls by assuming sparsity: only a subset of the predictors is included in the true model. In this context, the PDS-LASSO (Post-Double-Selection LASSO) methodology plays a crucial role. Ahrens, Hansen and Schaffer (2020) describe PDS-LASSO as a method that uses LASSO multiple times in a sequence to select controls. Initially, it performs a LASSO regression where the dependent variable is the original response variable itself. This step is designed to select the most relevant controls. Subsequently, PDS-LASSO runs a separate LASSO regression for each independent variable that is configured as non-penalized, treating each one as the dependent variable. This step results in the selection of additional controls. The final model incorporates the set of all controls selected through these stages. Upon completion of the control selection process, the original equation is run again, but this time it only includes the selected controls. In our study, we deviate from the traditional use of linear regression in the final step and instead employ an ordinal probit model. To summarize, our estimation methodology is a two-step process: initially, we use the PDS-LASSO model to select the relevant controls, and subsequently, we apply the ordered probit model to execute our final specification of airport satisfaction.[25]

---

[24] For studies on the relationship between passenger crowding and its spillover effects on passenger stress, see Cox, Houdmont & Griffiths (2006) and Mahudin, Cox & Griffiths (2011).

In summary, our study emphasizes the importance of considering psychosituational factors when investigating passenger satisfaction. These factors significantly impact how passengers respond and their level of satisfaction when participating in airport surveys or evaluations. For instance, individuals who have encountered stressful situations, such as dealing with heavy traffic or rushing to catch a flight, may exhibit higher levels of irritability, impatience, and anxiety. These emotional states can influence their overall evaluation of the airport experience and potentially affect their satisfaction ratings. Other relevant psychosituational factors in the context of airport satisfaction include airport layout and signage, availability of amenities, ease of security screening processes, and the overall airport atmosphere. These factors are generally applicable as they refer to the psychological and situational elements that shape passengers' emotions, behaviors, and perceptions throughout their airport journey, from before they arrive at the airport to their time spent there. While our study focuses on São Paulo/Guarulhos Airport, these factors can be applicable to other airport contexts as well. By including these psychosituational factors as control variables in our study, using variables and controls outlined in Equation (5) such as the DISSAT variables, we aim to account for some of their potential influence on passengers' perceptions and satisfaction. This approach allows us to isolate the specific impact of flight delays and other factors of interest on passenger satisfaction. It's important to note that these control variables do not encompass all psychosituational factors but serve as proxies or indicators for the broader range of psychological and situational aspects that passengers may experience.

---

[25] To address heteroskedasticity, the LASSO penalty loadings account for clustering of airport terminal interacted with survey date. See Oliveira et al. (2021) for a similar LASSO approach.

## 5. Estimation results

Table 2 presents our estimation results of the global airport passenger satisfaction model (APTSAT), dictated by Equations (1) and (5). It can be seen that the table presents the result of eight specifications. The main objective of these experiments was to estimate the ceteris paribus effect of DEL on APTSAT, considering different ways of controlling respondents' latent psychosituational traits. Therefore, we aimed to evaluate the effect of using the methodological proposal discussed in Section 3.4. In Column (1), we present a baseline model corresponding to Equation (1). In Column (2), we introduce a model where a potential sample selection bias regarding the surveyed passengers is addressed: we expanded the sampling of business travelers using a Synthetic Minority Oversampling Technique (SMOTE) approach.[26] As Franz et al. (2021) discuss, addressing selection bias in airport passenger surveys presents a complex challenge, as a lack of willingness to participate leads to a complete absence of information about these non-respondents. According to raw data from a survey conducted by the Ministry of Tourism at Brazilian airports in 2010, the non-response rate was approximately 27%.[27] Halse et al. (2022) argue that those who allocate time to answer a survey may typically have a lower opportunity cost of time than the average traveler. While the authors focus on the context of online surveys, it can be posited that, generally speaking, business travelers have a higher opportunity cost of time than leisure travelers. Consequently, we employ the SMOTE procedure to enhance the representation of business travelers in the sample and verify the robustness of our empirical results.[28]

In Column (3), we present a model where DEL is alternatively measured as a flight delay of 30 minutes or more, in accordance with the regulations of ANAC (Brazilian National Civil Aviation Agency). Columns (5) to (8) show our preferred results, given that they come from specifications in which all the controls for the unobservable related to the passenger's personality and emotional context are included.

First, we highlight the DEL variable estimation. In all specifications, the estimated marginal effect of this variable was negative and statistically significant. The results confirm the existence of a relationship between flight delays and passengers' overall satisfaction with airports. This result is intuitive and expected ex-ante. However, note that the estimated coefficient ranges from -0.0799 (Column 1) to -0.0703 (Column 4), indicating an absolute marginal effect drop of approximately 12.02%. This result is clearly due to the inclusion of controls for unobservable psychological traits of respondents (DISSAT), which are absent from the specification in Column (1) and are fully inserted in Column (4). When we include other psychological treatment variables such as TERMDIS and the pandemic variables (Column 5), this effect decreases to -0.0563 (-29.54%). Thus, we obtain evidence that a specification that does not consider passengers' psychosituational traits can cause a relevant overestimation of the absolute effect of flight delays on passenger satisfaction, probably because of the implied bias of omitted variables.

---

[26] Chawla et al (2002).

[27] Ministry of Tourism - MTUR and Foundation Institute of Economic Research - FIPE, International Tourism Demand Study - 2004-2010. Available at www.dadosefatos.turismo.gov.br.

[28] While these arguments suggest that, in principle, business passengers could significantly contribute to selection bias in passenger surveys, it is important to recognize that individual experiences and behaviors can vary greatly. Therefore, further empirical investigation is necessary to confirm such assumption.

It is important to note that although the estimated coefficients of the delay variable presented in Table 2 fall within the range of -0.0563 (Column 5) to -0.3423 (Column 6), which may initially be considered a small magnitude, the effect is still statistically significant. To further explore this discussion, we conducted a simulation of the predicted changes in passenger ratings based on the occurrence or absence of flight delays. Utilizing the results from Columns (5) and (6), our ordered probit models predicted that in the event of a flight delay, between 4.31% and 27.31% of the surveyed passengers would reduce their satisfaction ratings for the airport by at least 1 point on a scale of 1 to 10. This would result in a projected average decrease in satisfaction ranging from 0.85% (Column 5) to 5.31% (Column 6). As a balance of this analysis, we conclude that the estimated effect of delays on passenger ratings can be considered relatively small on average. However, in the case of internal-sourced delays, the effect can be more prominent and therefore should be closely monitored. Considering that in Brazil, airport performance below regulatory standards can result in high contractual penalties and a lower annual tariff adjustment, any quality performance falling short of passenger expectations could ultimately lead to significant revenue losses due to existing service quality regulations.

An interesting result in Columns (4) and (5) is the statistical significance of DISSAT (AIRLINE). Although the estimated coefficient of this variable is systematically lower than the coefficients of the other DISSAT variables, these results suggest that dissatisfaction arising from passengers' interaction with the airline may exert a negative spillover effect on satisfaction with the airport.[29] This result suggests that the airport should engage in partnerships with airlines to improve the perceived quality of its service, targeting the prevention of possible contamination of dissatisfaction caused by the airline in its own levels of satisfaction. As expected ex-ante, the other psychological factors controlling DISSAT and TERMDIS present negative and statistically significant coefficients. In particular, the statistical significance of TERMDIS also reveals a negative spillover effect on the surveyed passenger from other passengers on delayed flights at the time of the interview. This result indicates that psychosituational traits associated with a greater agglomeration of people at times of many delayed flights at the terminal are relevant in explaining passenger dissatisfaction during this period. In terms of the pandemic variables, only PANDEMIC (LATER) was found to be statistically significant, albeit at a 10% level. Interestingly, this variable has a positive coefficient, which may indicate a traveler's optimism in resuming their trips in the post-pandemic period. It is important to note that, more than the statistical significance of these pandemic variables and flight date controls, the key aspect of our methodology is the control they provide for unobservable factors, aiding in a more accurate and consistent estimation of the model.

---

[29] See Weber & Sparks (2004) for a similar spillover effect between partner firms.

In Column (6) of Table 2, we present the results of a two-step procedure to investigate the issue of service failure and attribution of blame, as discussed in Section 2.3. In Step I, we estimate the determinants of flight delays by considering internal and external factors. The results of this model can be found in Table 5 in the Appendix. In Step II, using the predictions from the flight delay model, we create variables representing delays of internal and external origins. These variables are included in Column (6), along with psychological traits. Our analysis reveals that delays of internal origin have a significant impact on passenger satisfaction, while delays of external origin are dropped by the PDS-LASSO penalty procedure. This suggests that passengers attribute blame to the airport for delays caused by internal factors, as these delays remain among the selected variables despite the penalty. These findings have important implications for quality management in the airline and airport industry, emphasizing the need to monitor passenger sensitivity to delays and address dissatisfaction when weather conditions do not affect airport operations.

Finally, we present the results from Columns (7) and (8) of Table 2. In these columns, we include interaction variables between DEL and the following boarding period dummy variables: BOARD (NOT), BOARD (CALL), and BOARD (NOW). These variables represent, respectively, a period before the flight when boarding is not yet open, a period when the boarding call is imminent, and a period when boarding was likely underway during the survey. The delineation of these periods is somewhat arbitrary, so we use two configurations. We use 40 minutes before the scheduled departure time as a reference for domestic flights, and 1 hour for international flights. These time thresholds are used to differentiate between BOARD (NOW) and BOARD (CALL). For the threshold that distinguishes between BOARD (NOT) and BOARD (CALL), we use the 75th percentile of time to flight at the time of the interview (Column 7) and the 90th percentile (Column 8).

**Table 2 – Estimation results: airport passenger satisfaction (APTSAT)**

| Variable | (1) | (2) | (3) | (4) | (5) | (6) | (7) | (8) |
|---|---|---|---|---|---|---|---|---|
| GENSILEN | 0.1399*** | 0.1370*** | 0.1379*** | 0.0960* | 0.0957* | 0.0939* | 0.0981* | 0.0964* |
| GENBOOM | 0.0295 | 0.0105 | 0.0291 | -0.0019 | -0.0139 | -0.0106 | -0.0130 | -0.0126 |
| GENMILLEN | 0.1942*** | 0.1906*** | 0.1943*** | 0.1369*** | 0.1292*** | 0.1291*** | 0.1325*** | 0.1324*** |
| GENZ | 0.3621*** | 0.3669*** | 0.3621*** | 0.3000*** | 0.2805*** | 0.2819*** | 0.2798*** | 0.2773*** |
| SCHLELEM | 0.7943*** | 0.8189*** | 0.7918*** | 0.6944*** | 0.7007*** | 0.6958*** | 0.7103*** | 0.7107*** |
| SCHLMIDD | 0.4893*** | 0.5061*** | 0.4893*** | 0.4163*** | 0.4197*** | 0.4133*** | 0.4316*** | 0.4307*** |
| SCHLHIGH | 0.3060*** | 0.3071*** | 0.3052*** | 0.2315*** | 0.2349*** | 0.2277*** | 0.2393*** | 0.2392*** |
| FIRSTTFLIER | 0.1281*** | 0.1241*** | 0.1279*** | 0.0834*** | 0.0785*** | 0.0785*** | 0.0770*** | 0.0779*** |
| FREQFLIER | -0.2166*** | -0.2297*** | -0.2169*** | -0.1547*** | -0.1519*** | -0.1508*** | -0.1520*** | -0.1511*** |
| LSRFLIER | 0.0467** | 0.0494*** | 0.0467** | 0.0364* | 0.0405** | 0.0416** | 0.0378* | 0.0370* |
| INTNLDEST | 0.0143 | 0.0350 | 0.0153 | -0.1074 | -0.0620 | -0.1146 | -0.0715 | -0.0689 |
| REDEYE | -0.0977** | -0.1102*** | -0.1016*** | -0.0457 | -0.0526 | -0.0392 | -0.0501 | -0.0466 |
| SMALLTERM | -0.0788 | -0.1149 | -0.0779 | -0.1089 | -0.0409 | -0.0000 | -0.0479 | -0.0542 |
| INTNLTERM | 0.1774** | 0.1908*** | 0.1799** | 0.3391*** | 0.3361*** | 0.3291*** | 0.3295*** | 0.3293*** |
| TERMDEN | -0.3218*** | -0.2987*** | -0.3260*** | -0.2488*** | -0.1943** | -0.1858** | -0.1884** | -0.1855** |
| JETBRIDGE | 0.0368 | 0.0492** | 0.0364 | 0.0880*** | 0.0866*** | 0.0835*** | 0.0890*** | 0.0912*** |
| SHOPS | 0.1753*** | 0.1535*** | 0.1753*** | 0.1290*** | 0.1327*** | 0.1328*** | 0.1314*** | 0.1321*** |
| FOOD | 0.4709*** | 0.4875*** | 0.4713*** | 0.3836*** | 0.3888*** | 0.3882*** | 0.3954*** | 0.3944*** |
| EXPENSIVE | -0.3053*** | -0.3171*** | -0.3046*** | -0.0548 | -0.0540 | -0.0549 | -0.0515 | -0.0517 |
| WIFI | 0.2359*** | 0.2494*** | 0.2358*** | 0.1137*** | 0.1062*** | 0.1078*** | 0.1086*** | 0.1083*** |
| DISSAT (CURBSID) | | | | -0.0404*** | -0.0411*** | -0.0414*** | -0.0411*** | -0.0408*** |
| DISSAT (CHECKIN) | | | | -0.0264*** | -0.0264*** | -0.0267*** | -0.0261*** | -0.0262*** |
| DISSAT (WAYFIND) | | | | -0.2152*** | -0.2171*** | -0.2168*** | -0.2171*** | -0.2168*** |
| DISSAT (WALKDST) | | | | -0.2646*** | -0.2639*** | -0.2636*** | -0.2642*** | -0.2639*** |
| DISSAT (FLTINFO) | | | | -0.1831*** | -0.1825*** | -0.1825*** | -0.1832*** | -0.1837*** |
| DISSAT (SECINSP) | | | | -0.1563*** | -0.1587*** | -0.1578*** | -0.1585*** | -0.1588*** |
| DISSAT (AIRLINE) | | | | -0.0167*** | -0.0158*** | -0.0158*** | -0.0160*** | -0.0160*** |
| TERMDIS | | | | | -0.1167** | -0.0597 | -0.1061** | -0.1173** |
| PANDEMIC (PRE) | | | | | -0.0188 | -0.0124 | -0.0206 | -0.0227 |
| PANDEMIC (EARLY) | | | | | -0.0146 | -0.0103 | 0.0036 | 0.0064 |
| PANDEMIC (LATER) | | | | | 0.1069* | 0.0957 | 0.1463** | 0.1075* |
| DEL | -0.0799*** | -0.0831*** | -0.0900*** | -0.0703*** | -0.0563** | | | |
| DEL (INT) | | | | | | -0.3423*** | | |
| DEL (EXT) | | | | | | (lasso drop) | | |
| DEL × BOARD (NOT) | | | | | | | -0.0686 | -0.0611 |
| DEL × BOARD (CALL) | | | | | | | -0.0645** | -0.0580** |
| DEL × BOARD (NOW) | | | | | | | -0.0097 | -0.0058 |
| Post Lasso Estimator | Ordered Probit PDS-LASSO | Ordered Probit PDS-LASSO | Ordered Probit PDS-LASSO | Ordered Probit PDS-LASSO | Ordered Probit PDS-LASSO | Ordered Probit PDS-LASSO | Ordered Probit PDS-LASSO | Ordered Probit PDS-LASSO |
| Clusters | AptTerm/Date | AptTerm/Date | AptTerm/Date | AptTerm/Date | AptTerm/Date | AptTerm/Date | AptTerm/Date | AptTerm/Date |
| Log-likelihood | -21,868 | -25,491 | -21,869 | -20,099 | -20,046 | -20,062 | -20,020 | -20,018 |
| AIC Statistic | 44,131 | 51,371 | 44,134 | 40,584 | 40,558 | 40,565 | 40,649 | 40,646 |
| BIC Statistic | 45,604 | 52,858 | 45,615 | 42,027 | 42,300 | 42,218 | 42,930 | 42,927 |
| Flight Date Controls | yes | yes | yes | yes | yes | yes | yes | yes |
| Airline Controls | yes | yes | yes | yes | yes | yes | yes | yes |
| Destination Controls | yes | yes | yes | yes | yes | yes | yes | yes |
| Time-to-Flight Controls | yes | yes | yes | yes | yes | yes | yes | yes |
| SMOTE Sampling | no | yes | no | no | no | no | no | no |
| Nr Observations | 13,071 | 15,158 | 13,071 | 13,071 | 13,071 | 13,071 | 13,071 | 13,071 |

*Notes: Estimation results produced by ordered probit regression with robust standard errors. Variable selection performed by a previous procedure using the post-double-selection LASSO-based methodology (PDS-LASSO) of Belloni et al. (2012, 2014a,b). LASSO penalty loadings account for the clustering of airport terminal and survey date. Flight date, airline, and destination control variables estimates omitted. Variables set as under LASSO penalization: DISSAT (all variables), TERMDIS, PANDEMIC, DEL (INT, EXT), and the proposed controls. In Column (2), SMOTE (Survey Minority Oversampling Technique) was employed to address class imbalance in survey data. In Column (3), DEL was computed considering delays of 30 minutes. Column (6) was estimated using a previous estimation of delay determinants (available in the Appendix). DEL (INT) and DEL (EXT) represent delays of internal and external origin, respectively. Columns (7) and (8) differ only in the definition of BOARD (NOT, CALL, NOW). Please refer to the variable description for more details. P-value representations: \*\*\*p<0.01, \*\*p<0.05, \*p<0.10.*

The results from Columns (7) and (8) of Table 2 highlight key elements in shaping passengers' perceptions of airport service quality. We found that only the interviews during the BOARD (CALL) period were statistically significant, indicating that this phase is when passengers gain crucial information regarding the risk/certainty of flight delays, affecting their satisfaction with the airport. Before this, the insignificance of DEL x BOARD (NOT) suggests that passengers lack sufficient information about their flight status and therefore, do not form any significant opinions. Beyond this period, during boarding, the insignificance of DEL x BOARD (NOW) suggests that passengers' expectations may be influenced by the airline's boarding management tactics, such as gate announcements about upcoming boarding events. These interactions, often devoid of details about potential delays, can instill a false sense of satisfaction, causing mood swings. However, airlines can mitigate dissatisfaction by providing updated flight information via text messages and offering services like food vouchers during extended delays.

The other results presented in Table 2 have a direct ceteris paribus interpretation. Regarding GENSILEN, GENBOOM, GENMILLEN, and GENZ, we find that passengers of the Silent, Millennial, and Z generations evaluate the airport better than passengers of the X generation (base case of the dummies) and Baby-Boomers (not significant coefficient). Regarding SCHLELEM, SCHLMIDD, and SCHLHIGH, we found that the higher the education level, the worse the passenger's evaluation. Regarding FIRSTTFLIER and FREQFLIER, frequent flier passengers are the most critical of the airport, probably because they have more travel experience and better knowledge of the reality of other airports. Regarding INTNLDEST, REDEYE, SMALLTERM, INTNLTERM, TERMDEN, and JETBRIDGE, we have evidence that the terminal dedicated to international operations (Terminal 3) of GRU and flights that their boarding is carried out through jetbridges present, on average, better evaluations. In addition, busy times are likely to naturally generate greater queues and crowding with the effect of decreasing satisfaction. Finally, regarding SHOPS, FOOD, EXPENSIVE, and WIFI, we find that positive evaluations of stores, restaurants, and the airport's Wi-Fi service have a positive effect on global passenger satisfaction and that the cost of products and services at the airport apparently does not change the overall passenger satisfaction. Despite the negative effect of EXPENSIVE observed in Columns (1) to (3), this effect dissipates in the subsequent columns, likely due to its correlation with passengers' psychological traits.

Regarding the experiments in Column (2) of Table 2, we perform the following steps for the SMOTE configuration: First, we identify the samples belonging to the minority class, which refers to respondents who are business travelers. Next, we select the nearest neighbor for each of these observations using the Euclidean distance in the model's variable space. We then calculate the difference between each observation and its corresponding neighbor for each variable. Afterward, we multiply this difference by a random number between 0 and 1 and add it to the original values. This process generates a new synthetic observation of a "business traveler" that is related to, but not identical to, the original one. We repeat this procedure to increase the participation of the business traveler segment from the original sample's 30.42% to 40% by introducing additional synthetic observations with the random difference procedure. Consequently, the sample size increases from 13,071 to 15,158 observations. We find that the results of the model using the extended data set in Column (2) are robust and similar to those in Column (1). To illustrate the impact of this procedure more effectively, Table 3 presents the results of the DEL variable for increasing SMOTE oversampling rates, ranging from 35% to 55%. Please note that we construct the results in Table 3 by extracting the mean and standard deviation of the DEL coefficient estimates from 250 random replications. In contrast, in Column (2) of Table 2, we perform a single replication to create the synthetic samples. It is worth mentioning that as the participation of business passengers increases, we observe a tendency for the DEL estimate to have higher absolute values, indicating that our estimate in Column (1) is conservative.

Table 3– Estimation results for DEL coefficient using increasing SMOTE oversampling rates (%)

| Variable | Statistic | Original Sample | SMOTE Sample | | | | |
|---|---|---|---|---|---|---|---|
| | | | 1 | 2 | 3 | 4 | 5 |
| BSNFLIER | Sample Mean | 30.42% | 35% | 40% | 45% | 50% | 55% |
| | Sample Size (0) Sample Size (1) | 9,095 3,976 | 9,095 4,897 | 9,095 6,063 | 9,095 7,441 | 9,095 9,095 | 9,095 11,116 |
| DEL | Mean Estimate % var. | -0.0781 - | -0.0795 1.9% | -0.0807 3.4% | -0.0820 5.0% | -0.0835 7.0% | -0.0852 9.1% |
| | Mean Std Error % var. | 0.0262 - | 0.0063 -75.8% | 0.0079 -69.6% | 0.0086 -67.1% | 0.0086 -67.1% | 0.0062 -76.1% |

*Notes: SMOTE (Survey Minority Oversampling Technique) was employed to address class imbalance in survey data (Chawla et al., 2002). Mean and standard deviation extracted from the average of 250 random replications of the procedure. Estimation results derived from a specification and estimator similar to those used in Table 2, Column (2).*

**Table 4–Estimation results: airport passenger satisfaction (APTSAT)–additional experiments**

| Variable | (1) | (2) | (3) | (4) |
|---|---|---|---|---|
| GENSILEN | 0.0952* | 0.0947* | 0.0949* | 0.0933* |
| GENBOOM | -0.0148 | -0.0129 | -0.0142 | -0.0140 |
| GENMILLEN | 0.1291*** | 0.1298*** | 0.1289*** | 0.1284*** |
| GENZ | 0.2808*** | 0.2821*** | 0.2800*** | 0.2787*** |
| SCHLELEM | 0.6993*** | 0.6965*** | 0.6998*** | 0.6990*** |
| SCHLMIDD | 0.4172*** | 0.4215*** | 0.4211*** | 0.4199*** |
| SCHLHIGH | 0.2341*** | 0.2337*** | 0.2353*** | 0.2349*** |
| FIRSTTFLIER | 0.1004*** | 0.0771*** | 0.0784*** | 0.0776*** |
| FREQFLIER | -0.1399*** | -0.1503*** | -0.1513*** | -0.1512*** |
| LSRFLIER | 0.0402* | 0.0406** | 0.0399* | 0.0385* |
| INTNLDEST | -0.0634 | -0.0627 | -0.0630 | -0.0432 |
| REDEYE | -0.0538 | -0.0547 | -0.0525 | -0.0524 |
| SMALLTERM | -0.0489 | -0.0409 | -0.0406 | -0.0406 |
| INTNLTERM | 0.3370*** | 0.3328*** | 0.3373*** | 0.3415*** |
| TERMDEN | -0.1972** | -0.1945** | -0.1936** | -0.1943** |
| JETBRIDGE | 0.0857*** | 0.0858*** | 0.0867*** | 0.0873*** |
| SHOPS | 0.1337*** | 0.1231*** | 0.1328*** | 0.1316*** |
| FOOD | 0.3880*** | 0.3664*** | 0.3893*** | 0.3883*** |
| EXPENSIVE | -0.0536 | -0.0549 | -0.0530 | -0.0535 |
| WIFI | 0.1067*** | 0.0816*** | 0.1063*** | 0.1053*** |
| DISSAT (CURBSID) | -0.0408*** | -0.0413*** | -0.0411*** | -0.0412*** |
| DISSAT (CHECKIN) | -0.0265*** | -0.0258*** | -0.0263*** | -0.0262*** |
| DISSAT (WAYFIND) | -0.2175*** | -0.2165*** | -0.2173*** | -0.2172*** |
| DISSAT (WALKDST) | -0.2642*** | -0.2625*** | -0.2642*** | -0.2640*** |
| DISSAT (FLTINFO) | -0.1823*** | -0.1824*** | -0.1826*** | -0.1824*** |
| DISSAT (SECINSP) | -0.1583*** | -0.1586*** | -0.1587*** | -0.1588*** |
| DISSAT (AIRLINE) | -0.0159*** | -0.0162*** | -0.0160*** | -0.0162*** |
| TERMDIS | -0.1133** | -0.1119** | -0.1236** | -0.1114** |
| PANDEMIC (PRE) | -0.0204 | -0.0206 | -0.0236 | -0.0202 |
| PANDEMIC (EARLY) | -0.0161 | -0.0187 | -0.0163 | -0.0151 |
| PANDEMIC (LATER) | 0.0849 | 0.1193** | 0.1043* | 0.1056* |
| DEL × FIRSTTFLIER | -0.1175** | | | |
| DEL × EXPERCDFLIER | 0.0125 | | | |
| DEL × FREQFLIER | -0.0599 | | | |
| DEL × FOOD (4/5 RATING) | | 0.1048** | | |
| DEL × SHOPS (4/5 RATING) | | 0.0476 | | |
| DEL × WIFI (4/5 RATING) | | 0.1518*** | | |
| DELDUR | | | -0.0948** | |
| DELDUR × LSRFLIER | | | | -0.1086* |
| DELDUR × BSNFLIER | | | | -0.1687** |
| DELDUR2 | | | 0.0290** | |
| DELDUR2 × LSRFLIER | | | | 0.0312* |
| DELDUR2 × BSNFLIER | | | | 0.0516** |
| Post Lasso | Ordered Probit | Ordered Probit | Ordered Probit | Ordered Probit |
| Estimator | PDS-LASSO | PDS-LASSO | PDS-LASSO | PDS-LASSO |
| Clusters | AptTerm/Date | AptTerm/Date | AptTerm/Date | AptTerm/Date |
| Log-likelihood | -20,043 | -20,037 | -20,046 | -20,043 |
| AIC Statistic | 40,616 | 40,607 | 40,619 | 40,619 |
| BIC Statistic | 42,597 | 42,604 | 42,594 | 42,616 |
| Flight Date Controls | yes | yes | yes | yes |
| Airline Controls | yes | yes | yes | yes |
| Destination Controls | yes | yes | yes | yes |
| Time-to-Flight Controls | yes | yes | yes | yes |
| Nr Observations | 13,071 | 13,071 | 13,071 | 13,071 |

*Notes: Estimation results produced by ordered probit regression with robust standard errors. Variable selection performed by a previous procedure using the post-double-selection LASSO-based methodology (PDS-LASSO) of Belloni et al. (2012, 2014a,b). LASSO penalty loadings account for the clustering of airport terminal, and survey date. Flight date, airline, and destination control variables estimates omitted. Variables set as under LASSO penalization: DISSAT (all variables), and the proposed controls. P-value representations: ***p<0.01, ** p<0.05, * p<0.10.*

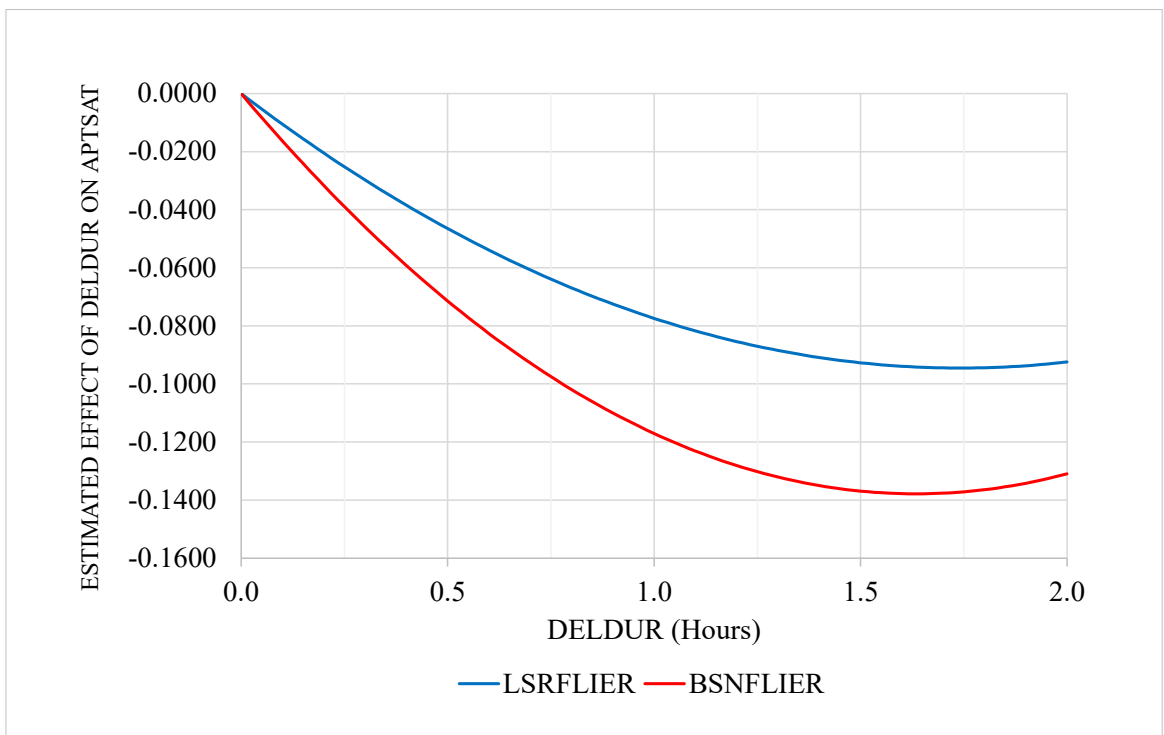

*Source: Estimation results from the interaction of DELDUR and DELDUR2 with LSRFLIER and BSNFLIER, as shown in Table 4, Column (4).*

**Figure 5–Effects of DELDUR on APTSAT–Differential impact on leisure and business travelers**

To examine the robustness of our preferred models in Table 3, we present the results of additional experiments in Table 4. Four specifications are presented: the interaction of DEL with passenger profile variables relating to travel experience (Column 1), the interaction of DEL with passenger experience variables pertaining to used airport services (Column 2), and the use of a continuous delay variable instead of a dummy variable (Columns 3 and 4).[30]

In Column (1) of Table 3, we can observe that first-time passengers are the most dissatisfied with delays (variable DEL × FIRSTFLIER), potentially due to their lack of experience with routine operational issues in air travel. These findings enable us to identify the customers most likely to be displeased with the airport, aiding in the development of strategies to alleviate the effects of delays.

In Column (2) of Table 3, we expand the analysis by introducing three additional flight delay interaction variables: DEL × FOOD (4/5 RATING), DEL × SHOPS (4/5 RATING), and DEL × WIFI (4/5 RATING). These variables represent the product of DEL with dummy variables that indicate the highest passenger satisfaction scores associated with airport food, retail stores, and Wi-Fi, respectively. The inclusion of these variables aims to examine the potential moderating effects of the quality of these items on passenger dissatisfaction resulting from flight delays. It is noteworthy that the DEL × FOOD (4/5 RATING) and DEL × WIFI (4/5 RATING) variables are statistically significant. This suggests that positive evaluations of the airport's restaurants and Wi-Fi have a meaningful attenuating effect on the discomfort experienced by passengers due to flight delays. These findings have significant implications for corporate policies and airport quality management strategies aimed at alleviating customer dissatisfaction caused by flight delays.

---

[30] Note that, if we use the Akaike Information Criterion (AIC) and the Bayesian Information Criterion (BIC) statistics, the models in the columns of Table 4 are relatively very similar, making it difficult to point out a preferred one. However, we emphasize that these experiments were designed solely for the purpose of conducting robustness checks in relation to our main results from Table 2. Furthermore, these AIC and BIC statistics from all columns of Table 4 are relatively close to our preferred specifications, that is, Columns (5) to (8) of Table 2.

Finally, in Columns (3) and (4), we present the results from substituting the DEL dummy variable with the continuous delay variable of more than 15 minutes from the scheduled departure time, labeled as DELDUR, along with its quadratic version, DELDUR2. To deepen the passenger segmentation analysis, we incorporate the interaction of DELDUR and DELDUR2 with the travel purpose dummy variables, LSRFLIER and BSNFLIER. From the results in Column (3), we can infer a potential quadratic relationship between DELDUR and APTSAT, as the coefficients of these variables are statistically significant at the 5% level. The negative sign of DELDUR and positive of DELDUR2 suggest a negative relationship that decreases in intensity as delay duration increases. This finding is intuitive, considering that in cases of flight delays, the legal/institutional framework surrounding passenger rights during such instances may come into play.[31] In Brazil, ANAC's Resolution No. 400/2016 stipulates that airlines provide assistance starting from a 1-hour delay, initially offering means of communication to the passenger (internet, phone calls, etc.), and providing food after 2 hours. The prospect of such assistance can offer comfort and recompense to passengers who experience substantial flight delays. This could reduce their anxiety, especially for leisure travelers as compared to business ones. To investigate this phenomenon, we developed Figure 5, which is based on the estimated coefficients of the interactions between DELDUR and LSRFLIER/BSNFLIER from Column (4) in Table 4. In the graph, it's possible to see the pattern of the estimated effect of DELDUR on APTSAT, which is negative and increases as the delay time increases. However, this pattern is attenuated to the extent that it stops falling around an hour and a half. Also note that the curve for business passengers is always below the curve for leisure passengers, which corroborates our initial expectations.

*5.1 Comparison with the literature*

Based on the results of our study, we find evidence of a relationship between flight delays and passengers' overall satisfaction with airports. Moreover, favorable evaluations of airport food/beverage services and Wi-Fi play a significant role in moderating this relationship. Additionally, we observe that passenger dissatisfaction resulting from interactions with the airline negatively impacts their satisfaction with the airport.

---

[31] We are grateful to an anonymous reviewer for suggesting this analysis.

Both Anderson, Bagget and Widener (2009) and our study agree that attributions of blame play a critical role in shaping customer satisfaction during service failures in the airline industry. Anderson, Bagget and Widener (2009) found that customer satisfaction is lower for all service failures, but the specific components of satisfaction differ between delayed and routine flights only when customers blame the service provider for the failure. Our study aligns with this, as we find that delays of internal origin significantly impact passenger satisfaction, indicating that passengers attribute blame to the airport for delays caused by internal factors. In contrast, delays of external origin are not considered as significant factors in our analysis, suggesting that passengers do not hold the service provider accountable for these delays.

When comparing our findings with those of Britto, Dresner, and Voltes (2012), we identify consistency between the two investigations. Both studies emphasize the correlation between flight delays and consumer welfare. Although our focus centers on passenger satisfaction while their research examines the impact on demand and consumer welfare, it is likely that these dimensions are interconnected, thereby validating the compatibility of the results.

The findings of Efthymiou et al. (2018) contradict the prevalent belief that British Airways passengers are irritated by delayed services at Heathrow Airport. In contrast, our study does not concentrate on a specific airline and does not directly address passengers' expectations regarding delays. However, we employ a high-dimensional econometrics approach to control for the emotional context of the passenger. Nevertheless, we recommend that future studies combine our approach with a more direct proxy for assessing passengers' expectations regarding delays.

Analyzing the study by Wen and Geng-qing Chi (2013), we note that both investigations acknowledge the significant influence of customer emotions on their service evaluations. While the authors focus on the impact of perceived justice and consumer emotions on satisfaction with service recovery, our study highlights the effect of flight delays on passengers' overall satisfaction with airports, particularly emphasizing the moderating role of favorable evaluations of specific airport services. Overall, our results align with the findings of the authors.

Contrasting our findings with those of Chow (2014), we encounter seemingly contradictory results. While our investigation concentrates on passengers' overall satisfaction with airports, Chow examines the relationship between customer complaints and their expectations of airline service quality. While our results indicate a decrease in passenger satisfaction with airports due to flight delays, the author's findings suggest that on-time performance does not significantly affect customer complaints. It is important to note that our study captures passengers' emotions shortly before boarding, whereas the author's study focuses on customer complaints. These methodological disparities may explain the apparent contradiction in the results, highlighting the importance of considering and isolating multiple aspects and contexts when evaluating customer satisfaction in the airline industry. Additionally, Chow (2015) identifies a correlation between poor weather conditions, such as increased rainfall or decreased temperature, and customer complaints. In contrast, our study decomposes delays into their internal and external causes, enabling a separate estimation of the weather's effect. Unlike the author's findings, our results do not indicate external delays as significant determinants of passenger satisfaction.

Examining the findings of Song, Guo, and Zhuang (2020), we observe strikingly similar results that support the relationship between flight delays and satisfaction in airports. However, our study delves deeper by dissecting this relationship into satisfaction with the airport and satisfaction with the airline as isolated factors.

In summary, when comparing our study's results with those of the various studies mentioned, we identify consistencies and complementary findings. We believe our research confirms several previously established conclusions in the literature while also expanding knowledge in the field by examining specific aspects of the relationship between flight delays. These aspects include the negative spillover effect of passenger interactions with airlines, the isolated effect of blame attribution, and the potential moderating effects of specific airport services. These discoveries hold vital implications for the airline industry, highlighting the necessity of effective flight delay management, addressing passenger expectations, and enhancing service quality to ensure customer satisfaction and economic efficiency.

Considering the extensive body of research in the Airport Service Quality (ASQ) field, we stress that our study's primary innovation lies in extending the van Praag and Ferrer-i-Carbonell (2008) approach. This extension is designed to capture potentially significant interactions between personality and the pre-interview and during-interview environment that might otherwise be reflected as residuals in the airport satisfaction econometric model. Failing to account for these latent psychosituational factors could lead to inadequately measured airport evaluations and the introduction of estimation biases. Our empirical contribution enhances alignment with Mowen's (2000) meta-theoretical model, which posits that interactions between psychological and situational factors influence human behavior. This alignment constitutes a novel development in the literature and has proven especially valuable in our study of passenger satisfaction response formation, helping mitigate the potential presence of confounding factors in our estimations.

Our empirical approach falls within the fields of Motivation Research, Consumer Psychology, and Happiness Economics. Recent scholarly literature has extensively explored individuals' satisfaction across various aspects of life within this domain, encompassing diverse disciplines. Notable recent examples include Bucciol and Zarri (2020), who examined the repercussions of exposure to traumatic events, and Darçın (2017), who delved into the impact of air pollution on subjective well-being. Additionally, Hartung, Spormann, Moshagen, and Wilhelm (2021) scrutinized structural variations in life satisfaction among distinct age cohorts, while Loibl, Haurin, Brown, and Moulton (2020) meticulously examined the relationship between reverse mortgage borrowing and domain-specific as well as overall life satisfaction among older individuals. Moreover, McDonald and Powdthavee (2022) quantified the costs associated with informal caregiving, and Rojas and Watkins-Fassler (2022) probed into the influence of religious practice on life satisfaction. Stephan, Demir, Lasch, Vossen, and Werner (2023) undertook a study on the psychological well-being of hybrid entrepreneurs, whereas Viñas-Bardolet, Guillen-Royo, and Torrent-Sellens (2020) explored the link between job characteristics and life satisfaction in the EU. Finally, Will and Renz (2022) investigated the correlation between homeownership and life satisfaction among individuals in debt. In the field of transportation, however, research remains somewhat limited. An example is the study conducted by Urban and Máca (2013), which examined the relationships between traffic noise, noise annoyance, and life satisfaction. We recommend considering the further application of concepts and approaches from the field of Happiness Economics when addressing the psychological aspects of transportation-related issues.

*5.2 Practical implications*

The empirical findings reveal several practical implications for stakeholders across the aviation industry. To gain a better understanding of passenger moods that can influence their behavior and attitudes, including their potential impact on airport consumption and concession revenues, the study emphasizes the crucial role of flight punctuality in determining overall satisfaction. It is important to acknowledge that negative experiences of passengers from the moment they arrive at the airport can have a cascading effect on their emotional context and state, potentially leading to expressions of discontent with the airport. In this regard, even factors like traffic conditions during the journey to the airport can have an effect that should be monitored by airport management.

Recognizing the detrimental impact of flight delays on the passenger experience, airports and airlines can prioritize coordinated strategies to minimize delays. This approach aims to improve not only operational efficiency but also to mitigate passenger dissatisfaction, as it can have negative implications for both stakeholders involved. Furthermore, the study underscores the significance of food/beverage concessions and Wi-Fi services in shaping passenger satisfaction, especially during flight delays. Airports can leverage this information to improve these amenities and ensure a positive passenger experience, ultimately enhancing their reputation and competitiveness. Airlines can also capitalize on the value of Wi-Fi connectivity by utilizing it to send timely updates and detailed information about delayed flights, providing comfort and alleviating passenger anxiety. In these cases, it is important to explain to passengers whether the causes of delays are internal or external to the aviation sector, to minimize the effects of potential blame attribution on their part.

From an airport management perspective, the research highlights the need to address passenger dissatisfaction stemming from interactions with airlines. The results suggest potential negative spillover effects, emphasizing the importance of collaboration between airports and airlines to deliver seamless and efficient services to passengers. Moreover, the study's findings on blame attribution behavior underscore the significance of clear communication and transparency during weather-related flight delays. Airport management can improve communication channels to help passengers understand factors beyond their control and mitigate the adverse effects of weather-related disruptions. These efforts should go beyond mere flight information updates and include actions that alleviate passenger stress during prolonged waiting periods.

In terms of public policy implications, the empirical results emphasize the role of service quality regulation. Brazilian aviation policymakers can utilize the findings to design and enforce regulations that monitor the quality of services provided by the privatized airports. This includes evaluating the effectiveness of measures taken to reduce flight delays, ensuring fair and transparent pricing of airport services, and encouraging investments in amenities such as food/beverage concessions and Wi-Fi. However, these policies should be implemented carefully to avoid disincentivizing good airport management practices that would be pursued even without regulations. By aligning policy interventions with the identified determinants of passenger satisfaction, regulators can promote a positive passenger experience and foster a competitive and customer-oriented aviation industry. It is important to note that these considerations are more applicable to the national reality of privatized airports with concession contracts with the Brazilian government than to other contexts worldwide.

## 6. Conclusion

This study develops an econometric methodology for generating and selecting control variables for unobservable factors of respondents' personality and emotional context in empirical passenger satisfaction models. This methodology aims to avoid omitted variable bias in satisfaction studies that use survey data. Based on a model of the determinants of passenger satisfaction from a survey conducted at São Paulo/GRU Airport, we investigated the association between flight delays and airport passenger satisfaction.

The results of this study indicate a causal relationship between flight delays and passengers' overall satisfaction with airports. We also found evidence that favorable evaluations of airport restaurants and Wi-Fi service have a relevant moderating effect on the relationship between delays and satisfaction. We find no similar evidence concerning retail outlets at airports. Additionally, our results suggest that passengers may project their dissatisfaction with the airline service onto their evaluations of the airport, which characterizes a negative spillover effect. We also point out a negative spillover effect related to the perception of greater crowding and stress at the airport terminal in situations of delays in many flights. From specific modeling for an empirical test related to the attribution of blame behavior for delays by passengers, we found that only delays of internal origin (flight management) are significant, but not flight delays due to external factors (climatic factors). Finally, we conclude that a specification that does not consider passengers' psychosituational traits can overestimate between 12 and 30% of the absolute effect of flight delays on passenger satisfaction.

Our study has a set of limitations from which we identify suggestions for future studies. Our methodology for controlling for respondents' psychosituational aspects partly addresses these issues; however, we recommend a future research design in which a Heckit model of selection bias is used. Also, considering that our model is designed to assess the impact of latent psychological characteristics, it would be beneficial to employ a structural equation modeling approach, which encompasses both the measurement model and path analysis. We also recommend in-depth modeling of attribution of blame, aiming to dissociate the internal factors genuinely coming from the airport from the internal factors coming from other entities co-responsible for delays, such as airline and air traffic control. Additionally, we recommend the adaptation, deepening, and application of our methodology for use in online post-trip surveys, which researchers and companies are increasingly using. Finally, we suggest that future studies investigate the possibility that the questionnaires applied in the surveys contain initial questions that map the personality and emotional context of the passenger more directly and that some passengers can be selected for a more in-depth study to identify specific elements of the psychological profile of professionals in the field.

Another limitation of our research relates to passengers in transit. Our methodological approach involved excluding passengers who indicated that they were on connecting flights. However, passengers who would have had subsequent connecting flights are included in our data. We believe that, in the case of flight delays, passengers who have landed at GRU seeking a connecting flight may be exposed to an emotional context of potentially missing their connections. Additionally, passengers whose first flight originates from GRU may also be exposed to a similar context regarding a possible missed connection at their destination airport. We believe that explicitly identifying these situations in the data would allow for the development of a model that captures the potentially amplified effect of flight delays on passenger satisfaction. We recommend that future studies delve further into this subject to gain a more comprehensive understanding.[32]

With the increasing need for quality management on the part of transport operators, our research contributes by generating implications for corporate policies for airports in the sense of promoting closer ties and partnerships with airlines, a greater focus on passengers who are possibly more sensitive to irritability generated by waiting due to delays, and greater attention to the service provided in circumstances of delays caused by factors internal to the industry.

---

[32] We would like to express our gratitude to an anonymous reviewer for suggesting this discussion.

# References

Ahrens, A., Hansen, C. B., & Schaffer, M. E. (2020). LASSOPACK: Model selection and prediction with regularized regression in Stata. The Stata Journal, 20(1), 176-235.

Anderson, E. W., Fornell, C., & Mazvancheryl, S. K. (2004). Customer satisfaction and shareholder value. Journal of marketing, 68(4), 172-185.

Anderson, S. W., Baggett, L. S., & Widener, S. K. (2009). The impact of service operations failures on customer satisfaction: evidence on how failures and their source affect what matters to customers. Manufacturing & Service Operations Management, 11(1), 52-69.

Antwi, C. O., Fan, C. J., Ihnatushchenko, N., Aboagye, M. O., & Xu, H. (2020). Does the nature of airport terminal service activities matter? Processing and non-processing service quality, passenger affective image and satisfaction. Journal of Air Transport Management, 89, 101869.

Arora, A., & Mathur, S. (2020). Effect of airline choice and temporality on flight delays: A case of domestic flights in India. Journal of Air Transport Management, 88, 101918.

Arora, S. D., & Mathur, S. (2020). Effect of airline choice and temporality on flight delays. Journal of Air Transport Management, 86, 101813.

Barakat, H., Yeniterzi, R., & Martin-Domingo, L. (2021). Applying deep learning models to twitter data to detect airport service quality. Journal of air transport management, 91, 102003.

Bellizzi, M. G., Eboli, L., & Mazzulla, G. (2020). Air transport service quality factors: a systematic literature review. Transportation Research Procedia, 45, 218-225.

Belloni, A., Chen, D., Chernozhukov, V., & Hansen, C. (2012). Sparse models and methods for optimal instruments with an application to eminent domain. Econometrica, 80(6), 2369-2429.

Belloni, A., Chernozhukov, V., & Hansen, C. (2014a). Inference on treatment effects after selection among high-dimensional controls. The Review of Economic Studies, 81(2), 608-650.

Belloni, A., Chernozhukov, V., & Hansen, C. (2014b). High-dimensional methods and inference on structural and treatment effects. Journal of Economic Perspectives, 28(2), 29-50.

Bezerra, G. C., & Gomes, C. F. (2020). Antecedents and consequences of passenger satisfaction with the airport. Journal of Air Transport Management, 83, 101766.

Bezerra, G. C., & Gomes, C. F. (2020). Antecedents and consequences of passenger satisfaction with the airport. Journal of Air Transport Management, 83, 101766.

Bucciol, A., & Zarri, L. (2020). Wounds that time can't heal: Life satisfaction and exposure to traumatic events. Journal of Economic Psychology, 76, 102241.

Burrieza-Galán, J., Jordá, R., Gregg, A., Ruiz, P., Rodríguez, R., Sala, M. Torres, J., García-Albertos, P., Ros, O., & Herranz, R. (2022). A methodology for understanding passenger flows combining mobile phone records and airport surveys: Application to Madrid-Barajas Airport after the COVID-19 outbreak. Journal of Air Transport Management, 100, 102163.

Cao, M., Li, L., & Zhang, Y. (2023). Developing a passenger-centered airport: A case study of Urumqi airport in Xinjiang, China. Journal of Air Transport Management, 108, 102363.

Cárdenas, M., Mejía, C., di Maro, V., Graham, C., & Lora, E. (2009). Vulnerabilities and subjective well-being. Paradox and perception: Measuring quality of life in Latin America, 118-57.

Cervone, D., & Pervin, L. A. (2019). Personality: Theory and research. John Wiley & Sons.

Chawla, N. V., Bowyer, K. W., Hall, L. O., & Kegelmeyer, W. P. (2002). SMOTE: synthetic minority over-sampling technique. Journal of Artificial Intelligence Research, 16, 321-357.

Chernozhukov, V., Hansen, C., & Spindler, M. (2015). Post-selection and post-regularization inference in linear models with many controls and instruments. American Economic Review, 105(5), 486-90.

Choi, S., & Mattila, A. S. (2008). Perceived controllability and service expectations: Influences on customer reactions following service failure. Journal of Business Research, 61(1), 24-30.

Chonsalasin, D., Jomnonkwao, S., & Ratanavaraha, V. (2021). Measurement model of passengers' expectations of airport service quality. International journal of transportation science and technology, 10(4), 342-352.

Cox, T., Houdmont, J., & Griffiths, A. (2006). Rail passenger crowding, stress, health and safety in Britain. Transportation Research Part A: Policy and Practice, 40(3), 244-258.

Darçın, M. (2017). How air pollution affects subjective well-being. Well-being and Quality of Life: Medical Perspective, 211.

Eboli, L., Bellizzi, M. G., & Mazzulla, G. (2022). A literature review of studies analysing air transport service quality from the passengers' point of view. Promet-Traffic & Transportation, 34(2), 253-269.

Efthymiou, M., Arvanitis, P. and Papatheodorou, A. (2016) Institutional Changes and Dynamics in the European Aviation Sector: Implications for Tourism. In Pappas N. & Bregoli, I. (eds) Global Dynamics in Travel, Tourism and Hospitality. Hershey, Pennsylvania: IGI Global, 41-57 (ISBN: 9781522502012).

Efthymiou, M., Arvanitis, S., & Papatheodorou, A. (2016). The systemic implications of flight delays and their effects on tourism growth in Europe. Journal of Air Transport Management, 55, 24-32.

Efthymiou, M., Arvanitis, S., Papatheodorou, A., & Mason, K. (2019). Does British Airways consistently meet or exceed customer expectations? Evidence from London Heathrow Airport. Journal of Air Transport Management, 78, 63-71.

Efthymiou, M., Njoya, E. T., Lo, P. L., Papatheodorou, A., and Randall, D. (2019) The Impact of Delays on Customers' Satisfaction: An Empirical Analysis of the British Airways On-Time Performance at Heathrow Airport. Journal of Aerospace Technology and Management, 11: eXX18.

Forrester, W. R., & Maute, M. F. (2001). The impact of relationship satisfaction on attributions, emotions, and behaviors following service failure. Journal of Applied Business Research (JABR), 17(1).

Franz, J. D., Holbert, H. T., Garrow, L. A., Gosling, G. D., Kamp, M. V., Harmon, L., & Ward, S. (2021). Guidebook for Conducting Airport User Surveys and Other Customer Research (No. ACRP Project 01-43).

Gayle, P. G., & Yimga, J. A. (2018). Air travel on-time performance and consumer behavior: Evidence from the US airline industry. Journal of Air Transport Management, 72, 68-80.

Gayle, P. G., & Yimga, J. O. (2018). How much do consumers really value air travel on-time performance, and to what extent are airlines motivated to improve their on-time performance?. Economics of transportation, 14, 31-41.

Halpern, N., & Mwesiumo, D. (2021). Airport service quality and passenger satisfaction: The impact of service failure on the likelihood of promoting an airport online. Research in Transportation Business & Management, 41, 100667.

Halpern, N., & Mwesiumo, D. (2021). Airport service quality and passenger satisfaction: The impact of service failure on the likelihood of promoting an airport online. Research in Transportation Business & Management, 41, 100667.

Halse, A. H., Flügel, S., Kouwenhoven, M., de Jong, G., Sundfør, H. B., Hulleberg, N., Jordbakke, G., & Lindhjem, H. (2022). A minute of your time: The impact of survey recruitment method and interview location on the value of travel time. Transportation, 1-32.

Harding, E. J., Paul, E. S., & Mendl, M. (2004). Cognitive bias and affective state. Nature, 427(6972), 312-312.

Harris, K. E., Mohr, L. A., & Bernhardt, K. L. (2006). Online service failure, consumer attributions and expectations. Journal of Services Marketing, 20(7), 453-458.

Hartung, J., Spormann, S. S., Moshagen, M., & Wilhelm, O. (2021). Structural differences in life satisfaction in a US adult sample across age. Journal of personality, 89(6), 1232–1251.

He, L., Yang, D., & Li, J. (2021). Improving the Service Quality of Public Transit with Exclusive Bus Lanes: A Perspective from Passenger Satisfaction. Journal of Advanced Transportation, 2021.

Hult, G. T. M., Sharma, P. N., Morgeson III, F. V., & Zhang, Y. (2019). Antecedents and consequences of customer satisfaction: do they differ across online and offline purchases?. Journal of Retailing, 95(1), 10-23.

Iversen, S., Kupfermann, I., & Kandel, E. R. (2000). Emotional states and feelings. Principles of neural science, 4, 982-997.

Jiang, H., & Ren, X. (2019). Model of passenger behavior choice under flight delay based on dynamic reference point. Journal of Air Transport Management, 75, 51-60.

Jiang, R., & Ren, F. (2019). Understanding air passengers' decision-making behavior under flight delay: A dynamic reference point perspective. Journal of Air Transport Management, 81, 101723.

Laisak, A. H., Rosli, A., & Sa'adi, N. (2021). The Effect of Service Quality on Customers' Satisfaction of Inter-District Public Bus Companies in the Central Region of Sarawak, Malaysia. International Journal of Marketing Studies, 13(2), 1-53.

Li, L., Mao, Y., Wang, Y., & Ma, Z. (2022). How has airport service quality changed in the context of COVID-19: A data-driven crowdsourcing approach based on sentiment analysis. Journal of Air Transport Management, 105, 102298.

Loibl, C., Haurin, D. R., Brown, J. K., & Moulton, S. (2020). The relationship between reverse mortgage borrowing, domain and life satisfaction. The Journals of Gerontology: Series B, 75(4), 869–878.

Lopez-Valpuesta, L., & Casas-Albala, D. (2023). Has passenger satisfaction at airports changed with the onset of COVID-19? The case of Seville Airport (Spain). Journal of Air Transport Management, 108, 102361.

MacLeod, A. K., & Byrne, A. (1996). Anxiety, depression, and the anticipation of future positive and negative experiences. Journal of abnormal psychology, 105(2), 286.

Mahudin, N. M., Cox, T., & Griffiths, A. (2011). Modeling the spillover effects of rail passenger crowding on individual well being and organisational behaviour. WIT Transactions on the built environment, 116, 227-238.

Mainardes, E. W., de Melo, R. F. S., & Moreira, N. C. (2021). Effects of airport service quality on the corporate image of airports. Research in Transportation Business & Management, 41, 100668.

Martin-Domingo, L., Martín, J. C., & Mandsberg, G. (2019). Social media as a resource for sentiment analysis of Airport Service Quality (ASQ). Journal of Air Transport Management, 78, 106-115.

McDonald, R., & Powdthavee, N. (2022). The shadow prices of voluntary caregiving: Using well-being panel data to estimate the cost of informal care. Journal of Benefit-Cost Analysis.

Monsuur, F., Enoch, M., Quddus, M., & Meek, S. (2021). Modeling the impact of rail delays on passenger satisfaction. Transportation Research Part A: Policy and Practice, 152, 19-35.

Mowen, J. C. (2000). The 3M model of motivation and personality: Theory and empirical applications to consumer behavior. Springer Science & Business Media.

Nikbin, D., Ismail, I., Marimuthu, M., & Abu-Jarad, I. Y. (2011). The impact of firm reputation on customers' responses to service failure: the role of failure attributions. Business Strategy Series, 12(1), 19-29.

Noyan, F., & Şimşek, G. G. (2014). The antecedents of customer loyalty. Procedia-Social and Behavioral Sciences, 109, 1220-1224.

Oliveira, A. V. M., Narcizo, R. R., Caliari, T., Morales, M. A., & Prado, R. (2021). Estimating fuel-efficiency while accounting for dynamic fleet management: Testing the effects of fuel price signals and fleet rollover. Transportation Research Part D: Transport and Environment, 95, 102820.

Pagliari, R., & Graham, A. (2019). An exploratory analysis of the effects of ownership change on airport competition. Transport Policy, 78, 76-85.

Prentice, C., & Kadan, M. (2019). The role of airport service quality in airport and destination choice. Journal of Retailing and Consumer Services, 47, 40-48.

Rocha, P. M., Costa, H. G., & Silva, G. B. (2022). Gaps, trends and challenges in assessing quality of service at airport terminals: a systematic review and bibliometric analysis. Sustainability, 14(7), 3796.

Rojas, M., & Watkins-Fassler, K. (2022). Religious practice and life satisfaction: A Domains-of-life Approach. Journal of Happiness Studies, 23(5), 2349–2369.

Schul, P., & Crompton, J. L. (1983). Search behavior of international vacationers: Travel-specific lifestyle and sociodemographic variables. Journal of Travel Research, 22(2), 25-30.

Seetanah, B., Teeroovengadum, V., & Nunkoo, R. (2020). Destination satisfaction and revisit intention of tourists: does the quality of airport services matter?. Journal of Hospitality & Tourism Research, 44(1), 134-148.

Smith, T. A. (2020). The role of customer personality in satisfaction, attitude-to-brand and loyalty in mobile services. Spanish Journal of Marketing-ESIC.

Song, C., Guo, J., & Zhuang, J. (2020). Analyzing passengers' emotions following flight delays: a 2011–2019 case study on SKYTRAX comments. Journal of Air Transport Management, 89, 101903.

Song, L., Guo, Z., & Zhuang, W. (2020). Understanding air passengers' emotions in response to flight delays: Evidence from SKYTRAX online comments. Journal of Air Transport Management, 89, 101955.

Stephan, M., Demir, C., Lasch, F., Vossen, A., & Werner, A. (2023). Psychological well-being of hybrid entrepreneurs: A replication and extension study using German panel data. Journal of Business Venturing Insights, 20, e00419.

Tibshirani, R. (1996). Regression shrinkage and selection via the Lasso. Journal of the Royal Statistical Society: Series B (Methodological), 58(1), 267-288.

Tseng, C. C. (2020). An IPA-Kano model for classifying and diagnosing airport service attributes. Research in Transportation Business & Management, 37, 100499.

Urban, J., & Máca, V. (2013). Linking traffic noise, noise annoyance and life satisfaction: A case study. International journal of environmental research and public health, 10(5), 1895–1915.

Usman, A., Azis, Y., Harsanto, B., & Azis, A. M. (2022). Airport service quality dimension and measurement: a systematic literature review and future research agenda. International Journal of Quality & Reliability Management, 39(10), 2302-2322.

van Praag, B., & Ferrer-i-Carbonell, A. (2008). Happiness quantified: A satisfaction calculus approach. OUP Oxford, revised edition.

Viñas-Bardolet, C., Guillen-Royo, M., & Torrent-Sellens, J. (2020). Job characteristics and life satisfaction in the EU: A domains-of-life approach. Applied Research in Quality of Life, 15, 1069–1098.

Weber, K., & Sparks, B. (2004). Consumer attributions and behavioral responses to service failures in strategic airline alliance settings. Journal of Air Transport Management, 10(5), 361-367.

Wen, B., & Chi, C. G. Q. (2013). Examine the cognitive and affective antecedents to service recovery satisfaction: A field study of delayed airline passengers. International Journal of Contemporary Hospitality Management, 25(3), 306-327.

Will, S., & Renz, T. (2022). In debt but still happy? Examining the relationship between homeownership and life satisfaction. SOEPpapers on Multidisciplinary Panel Data Research.

Yimga, J. (2020). Price and marginal cost effects of on-time performance: Evidence from the US airline industry. Journal of Air Transport Management, 84, 101769.

Yimga, J. A. (2020). The impact of on-time performance on price and marginal cost: Evidence from the US airline industry. Journal of Air Transport Management, 87, 101820.

Yimga, J. A., & Gorjidooz, J. (2019). The impact of schedule padding on airline choice behavior: A discrete choice demand model for air travel. Journal of Air Transport Management, 75, 164-174.

Yimga, J., & Gorjidooz, J. (2019). Airline schedule padding and consumer choice behavior. Journal of Air Transport Management, 78, 71-79.

Zou, H., & Hastie, T. (2005). Regularization and variable selection via the elastic net. Journal of the Royal Statistical Society: Series B (Statistical Methodology), 67(2), 301-320.

**Appendix A. Data sources**

- Passenger Satisfaction Survey ("PS Survey/ANAC") of São Paulo/Guarulhos – Governor André Franco Montoro International Airport. See www.nectar.ita.br/avstats/studies_reports.html for a link to the original website in Portuguese.

- Air Transport Statistical Database, Microdata ("ATSD-M/ANAC"): detailed air transport supply and demand information disaggregated at flight level (flight/airport-pair/airline/month). National Civil Aviation Agency (ANAC). See www.nectar.ita.br/avstats/anac_statdata.html for a description and a link to the original website in Portuguese.

- Active Scheduled Flight Historical Data Series ("VRA/ANAC"): detailed flight operations information containing flight numbers, departure and arrival times, delays, and cancellations. National Civil Aviation Agency (ANAC). See www.nectar.ita.br/avstats/anac_vra.html for a description and a link to the original website in Portuguese.

- Airport Capacity (Runway) Declaration of Guarulhos Airport – São Paulo/ANAC. Summer and Winter Seasons (2018 to 2021). Available on National Civil Aviation Agency (ANAC)'s http://www.gov.br/anac website www.gov.br/anac. http://www.gov.br/anac

- Iowa State University Mesonet ("IEM/Iowa State University"). The IEM's ASOS-AWOS-METAR Data provides automated airport weather observations worldwide. We collected IEM data on cloud ceilings concerning sky coverage and sky altitude in feet, visibility in miles, wind gusts in knots, thunderstorms, and hail events. Website: mesonet.agron.iastate.edu/ASOS

**Appendix B. Description of model variables**

- AIRCSIZE is the size of the aircraft, measured by the number of seats (divided by 100). Larger aircraft tend to generate greater logistical complexity for airlines in terms of boarding passengers and baggage. Sources: PS Survey/ANAC (flight number) and ATSD-M/ANAC (aircraft seat number).

- AIRL is a set of control variables (dummies) designed to accommodate the specific characteristics of each airline in the study; constructed using authors' classifications.

- APTSAT is the variable representing the global satisfaction of departing passengers with the airport, measured using a 10-point Likert scale in interviews at GRU Airport. Original question from the survey questionnaire: "*In general, I am satisfied with this airport (rate from 1 to 10).*".

It is important to note that this variable is extracted exclusively from passengers who are awaiting their departure and have not made a flight connection. Source: PS Survey/ANAC.

- BOARD (CALL) a period when the boarding call is imminent during the survey. See BOARD (NOW) for details. Sources: PS Survey/ANAC (interview time and flight number) and VRA/ANAC (flight time); constructed using authors' calculations and classifications.

- BOARD (NOT) is a dummy variable to account for surveys conducted during a period when boarding is not yet open. See BOARD (NOW) for details. Sources: PS Survey/ANAC (interview time and flight number) and VRA/ANAC (flight time); constructed using authors' calculations and classifications.

- BOARD (NOW) is a dummy variable to account for a period when boarding was likely underway during the survey. We use 40 minutes before the scheduled departure time as a reference for domestic flights, and 1 hour for international flights. These time thresholds are used to differentiate between BOARD (NOW) and BOARD (CALL). For the threshold that distinguishes between BOARD (NOT) and BOARD (CALL), we use the 75th percentile of time to flight at the time of the interview (Column 7) and the 90th percentile (Column 8). Sources: PS Survey/ANAC (interview time and flight number) and VRA/ANAC (flight time); constructed using authors' calculations and classifications.

- BSNFLIER is a dummy variable that takes the value of 1 if the passenger is traveling for business-related purposes. This variable aims to control for the effects on the average satisfaction among business travelers. Source: PS Survey/ANAC.

- BUSYDAY is the total number of passengers handled by the respective airport terminal (divided by ten thousand) on the day of a passenger's flight. Sources: PS Survey/ANAC (place of interview and flight number) and ATSD-M/ANAC (day of flight and number of daily passengers in the terminal); constructed using authors' calculations.

- BUSYHOUR is the total number of passengers handled by the respective airport terminal (divided by one thousand) during the full hour when the interviewed passenger's flight was scheduled. Sources: PS Survey/ANAC (place of interview and flight number) and ATSD-M/ANAC (day and scheduled time of flight and number of passengers per hour in the terminal); constructed using authors' calculations.

- CARGO is the total amount of paid cargo, free cargo, and mail (divided by 10) carried by passenger flight aircraft. Sources: PS Survey/ANAC (interview location and flight number) and ATSD-M/ANAC (total kilos loaded on the flight aircraft); constructed using authors' calculations.

- CASCAD (ARR) is the average proportion of delayed flights that landed at GRU Airport three hours before the interviewed passenger's scheduled flight time. We define a late landing as one made 15 min after the scheduled time. Sources: PS Survey/ANAC (flight number) and VRA/ANAC (flight time, total scheduled landings, and delayed landings in the last three hours); constructed using authors' calculations and classifications.

- CASCAD (DEP) is the average proportion of delayed flights that departed from GRU Airport three hours prior to the scheduled flight time of the interviewed passengers. We define a delayed takeoff as one that takes place 15 min after the scheduled time. Sources: PS Survey/ANAC (flight number) and VRA/ANAC (flight time, total scheduled departures, and delayed departures in the last three hours); constructed using authors' calculations and classifications.

- DATE is a set of control variables (dummies) designed to to capture the unique attributes of each survey date within the study; constructed using authors' classifications.

- DEL is a dummy variable that takes the value of 1 if the interviewed passenger's flight took off more than 15 min after the scheduled takeoff time from GRU Airport. Sources: PS Survey/ANAC (flight number) and VRA/ANAC (scheduled and actual departure times); constructed using authors' calculations.

- DEL (INT) is a proxy for the delays of internal origin. We built on the classification of Anderson et al. (2008). These authors considered different sources of delays, proposing a distinction between flight delays of internal origins (i.e., air transport industry-related) and flight delays of external origins (i.e., weather-related). To include elements of attribution of blame by passengers for flight delays in our modeling, we propose the following two-step procedure: <u>Step I</u>: Using a previously estimated econometric model, we estimate the determinants of flight delays (DEL), considering the covariate factors related to internal and external delays, as defined by Anderson et al. (2008). <u>Step II</u>: Using the predictions from the flight delay model estimated in Step I, we create two variables: DEL (INT) and DEL (EXT), which are the predictions of the probability of delays of internal origin and external origin, respectively. The covariates included in the econometric model of Step I are: WEATHER (ORG), WEATHER (DST), SMALLTERM, INTNLTERM, JETBRIDGE, PRCONNECT, LOADFAC, AIRCSIZE, CARGO, DISTANCE, BUSYDAY, BUSYHOUR, SECINSPTIME, RUNWAYCONG, RUNWAYDIS, CASCAD (DEP), CASCAD (ARR), PANDEMIC (EARLY), and PANDEMIC (LATER); SECINSPTIME was constructed using a procedure similar to DISSAT (AIRLINE), utilizing evaluations related to the security inspection time at the airport; constructed using authors' calculations.

- DEL (EXT) is a proxy for the delays of external origin. Refer to DEL (INT) for details on how this variable was constructed.

- DELDUR is a continuous variable that measures the extent of flight delays exceeding 15 minutes beyond the originally scheduled departure time. This variable is measured in hours of flight delays. Source: PS Survey/ANAC (flight number) and VRA/ANAC (scheduled and actual departure times); constructed using authors' calculations and classifications.

- DELDUR2 is the quadratic version of DELDUR. Source: PS Survey/ANAC (flight number) and VRA/ANAC (scheduled and actual departure times); constructed using authors' calculations and classifications.

- DEST is a set of control variables (dummies) to account for the idiosyncrasies of flight destinations within the study; constructed using authors' classifications.

- DISSAT (AIRLINE) is one of the control variables for unobservable psychological factors referring to passengers' psychological traits at the time of the interview. It is a proxy for the passenger's level of dissatisfaction with the airline's services in relation to the average dissatisfaction of the passengers interviewed at the time of the interview. To calculate the average, all passengers interviewed during the same full hour of the month of the interview were considered. The 5-point Likert scale applied in the PS Survey/ANAC questionnaire was used to calculate dissatisfaction. Missing comments were filled out with zero values on the original scale. The original question in the questionnaire was, '*How do you rate the service and courtesy of the check-in staff?*' The level of dissatisfaction was calculated as the difference between the maximum possible value (5) and the evaluation value. Source: PS Survey/ANAC; constructed using authors' calculations.

- DISSAT (CHECKIN) is one of the control variables for unobservable psychological factors of passengers' unobservable psychological traits at the time of the interview. This variable aims to control passenger dissatisfaction with check-in time at the airport in relation to the dissatisfaction of other passengers with the same item. See the details of its calculation in the description of the variable DISSAT (AIRLINE), which presents a similar calculation. The original question from the survey questionnaire: "*How do you rate the waiting time for check-in at the airport?*" Source: PS Survey/ANAC; constructed using authors' calculations.

- DISSAT (CURBSID) is one of the control variables for unobservable psychological factors of passengers' unobservable psychological traits at the time of the interview. This variable aims to control passenger dissatisfaction with the curb at the airport and possibly their state of dissatisfaction with urban traffic congestion during and until arrival at the same location in relation to the dissatisfaction of other passengers with this item. See the details of its calculation in the description of the variable DISSAT (AIRLINE), which presents a similar calculation. The original question from the survey questionnaire: "*How do you rate the ease of entering or leaving the vehicle on the access road next to the terminal entrance (curb)?*" Source: PS Survey/ANAC; constructed using authors' calculations.

- DISSAT (FLTINFO) is one of the control variables for unobservable psychological factors of passengers' unobservable psychological traits at the time of the interview. This variable aims to control passenger dissatisfaction when obtaining flight information inside the airport in relation to the dissatisfaction of other passengers with this item. See the details of its calculation in the description of the variable DISSAT (AIRLINE), which presents a similar calculation. The original question from the survey questionnaire: "*How do you rate the availability of flight information?*" Source: PS Survey/ANAC; constructed using authors' calculations.

- DISSAT (SECINSP) is one of the control variables for unobservable psychological factors of passengers' unobservable psychological traits at the time of the interview. This variable aims to control passenger dissatisfaction with the security inspection time for access to the airport's boarding area in relation to other passengers' dissatisfaction with this item. See the details of its calculation in the description of the variable DISSAT (AIRLINE), which presents a similar calculation. The original question from the survey questionnaire: "*How do you evaluate the waiting time in the security inspection queue?*" Source: PS Survey/ANAC; constructed using authors' calculations.

- DISSAT (WALKDST) is one of the control variables for unobservable psychological factors of passengers' unobservable psychological traits at the time of the interview. This variable aims to control passenger dissatisfaction with the distance traveled to their boarding gate in relation to the dissatisfaction of other passengers with this item. See the details of its calculation in the description of the variable DISSAT (AIRLINE), which presents a similar calculation. The original question from the survey questionnaire: "*How do you rate the distance walked in the passenger terminal?*" Source: PS Survey/ANAC; constructed using authors' calculations.

- DISSAT (WAYFIND) is one of the control variables for unobservable psychological factors of passengers' unobservable psychological traits at the time of the interview. This variable aims to control passenger dissatisfaction with wayfinding at the airport in relation to the dissatisfaction of other passengers with this item. See the details of its calculation in the description of the variable DISSAT (AIRLINE), which presents a similar calculation. The original question from the survey questionnaire: "*How do you rate the ease of finding your way in the terminal?*" Source: PS Survey/ANAC; constructed using authors' calculations.

- DISTANCE is the geodesic distance between GRU Airport and the respective destination airport of the interviewed passenger's flight. The Vincenty distance was considered an ellipsoidal model of the Earth. The values for this variable are in miles and divided by a thousand. Sources: PS Survey/ANAC (flight number) and ANAC (airport latitude and longitude); constructed using authors' calculations.

- EXPENSIVE measures the dissatisfaction of the interviewed passengers with the prices of restaurants and shops at the airport. A 5-point Likert scale was applied to the PS Survey/ANAC survey questionnaire to calculate the variable. The comments missing for this item in the survey were filled with a null value. The original questions in the questionnaire were "*How do you evaluate the price-quality ratio of restaurants?*" and "*How do you evaluate the price-quality ratio of the stores?*" The level of dissatisfaction was calculated as the difference between the maximum value (5) and the average value of restaurant and store reviews. The variable was rescaled to values between 0 and 1. Source: PS Survey/ANAC; constructed using authors' calculations.

- EXPERCDFLIER is a dummy variable that takes the value of 1 if the passenger has made one or two boardings at GRU Airport in the year prior to the interview. The original question in the survey questionnaire was, "*How many shipments have you made in the last 12 months?*". These passengers were the reference cases for the FIRSTTFLIER and FREQFLIER variables. Source: PS Survey/ANAC; constructed using authors' calculations and classifications.

- FIRSTTFLIER is a dummy variable that takes the value of 1 if the passenger has not boarded GRU Airport in the year prior to the interview. This variable aims to control for the effects on the average satisfaction of passengers who traveled by plane for the first time and passengers who possibly had not traveled for more than 12 months. The original question in the survey questionnaire was, "*How many shipments have you made in the last 12 months?*". As a reference case, this binary variable includes passengers who boarded 1 or 2 times at GRU in the year prior to the survey. See FREQFLIER and EXPERCDFLIER. Source: PS Survey/ANAC; constructed using authors' calculations and classifications.

- FOOD is a variable that measures the interviewed passengers' satisfaction with the quality and variety of food and beverage services available at the airport. A 5-point Likert scale was applied to the PS Survey/ANAC survey questionnaire to calculate the variable. The comments missing for this item in the survey were filled with a null value. The average responses regarding the quality and variety of services were recorded. The variable was rescaled to values between 0 and 1. The original question in the questionnaire was "*How do you rate the quality of restaurants/food facilities?*" and "*How do you rate the variety of restaurants/eating facilities?*". Source: PS Survey/ANAC; constructed using authors' calculations.

- FOOD (4/5 RATING) is a dummy variable that takes the value 1 if the passenger has rated the survey questionnaire items in the PS Survey/ANAC, which referred to the quality and variety of food and beverage services at the airport with a minimum of 4 points out of 5 (represented by "Good" or "Very Good"). The question in the questionnaire was "*How do you rate the quality of restaurants/food facilities?*" and "*How do you rate the variety of restaurants/eating facilities?*" Source: PS Survey/ANAC; constructed using authors' calculations.

- FREQFLIER is a dummy variable that takes the value of 1 if the passenger has made more than two boardings at GRU Airport in the year prior to the interview. This variable aims to control for the effects on the average satisfaction of frequent travelers. The original question in the survey questionnaire was, "*How many shipments have you made in the last 12 months?*" As a reference case, this binary variable has passengers who boarded 1 or 2 times in the GRU in the year prior to the survey. See also FIRSTTFLIER and EXPERCDFLIER for further details. Source: PS Survey/ANA; constructed using authors' calculations and classifications.

- GENBOOM is a dummy variable that takes the value of 1 if the interviewed passenger has been classified as belonging to the Baby-Boomer Generation, defined as people born between 1946 and 1964. The original question in the survey questionnaire asked the passengers to identify their age among the following alternatives: "(*1) up to 21 years; (2) 22 to 25 years; (3) 26 to 34 years; (4) 35 to 44 years; (5) 45 to 54 years; (6) 55 to 64 years; (7) 65 to 75 years; (8) 76 years or more*". This binary variable has passengers classified as belonging to generation X as the reference case. See also GENMILLEN, GENSILEN, and GENZ. Source: PS Survey/ANAC; constructed using authors' calculations and classifications.

- GENMILLEN is a dummy variable that takes the value of 1 if the interviewed passenger has been classified as belonging to the Millennial Generation, defined as people born between 1977 and 1995. See details in GENBOOM. Source: PS Survey/ANAC; constructed using authors' calculations and classifications.

- GENSILEN is a dummy variable that takes the value of 1 if the interviewed passenger has been classified as belonging to the Silent Generation, defined as people born before 1945. See details in GENBOOM. Source: PS Survey/ANAC; constructed using authors' calculations and classifications.

- GENX is a dummy variable that takes the value of 1 if the interviewed passenger has been classified as belonging to Generation X, defined as people born between 1965 and 1976. See details in GENBOOM. Source: PS Survey/ANAC; constructed using authors' calculations and classifications.

- GENZ is a dummy variable that takes the value of 1 if the interviewed passenger has been classified as belonging to Generation Z, defined as people born between 1996 and 2015. See details in GENBOOM. Source: PS Survey/ANAC; constructed using authors' calculations and classifications.

- INTNLDEST is a dummy variable that takes the value of 1 for interviewed passengers whose final destination is located abroad. The reference case for this binary variable is that of passengers with a final domestic destination. Source: PS Survey/ANAC (final trip destination).

- INTNLTERM is a dummy variable that takes the value of 1 for passengers interviewed near the boarding gates of Terminal 3 at GRU Airport. Terminal 3 is the newest and most modern terminal dedicated to international flights. The reference case for this binary variable is Terminal 2 (older and with domestic hybrid operations and international operations in general with destinations in Latin America). See also SMALLTERM. Source: PS Survey/ANAC (place of interview).

- JETBRIDGE is a dummy variable that takes the value of 1 for passengers interviewed near boarding gates that have jetbridges at GRU Airport. The reference case for this binary variable is the boarding gates, whose boarding is carried out by bus to a remote point of the operations yard. Source: PS Survey/ANAC (place of interview), with its classification based on internal location maps available on the airport website.

- LOADFAC is the proportion of seats occupied by passengers on the aircraft assigned to the flights of the interviewed passengers. Sources: PS Survey/ANAC (flight number) and ATSD-M/ANAC (number of passengers and seats in the aircraft).

- LSRFLIER is a dummy variable that takes the value of 1 if the passenger is traveling for leisure purposes. This variable aims to control for the effects on the average satisfaction among leisure travelers. For LSRFLIER, the reference category is passengers traveling for business purposes (BSNFLIER) and other travel motives. Source: PS Survey/ANAC.

- PANDEMIC (EARLY) is a dummy variable that takes the value of 1 for passengers interviewed during the initial period of the COVID-19 pandemic in March 2020. No interviews were conducted between April 2020 and December 2020. The reference case for this binary variable is the year 2018. See PANDEMIC (PRE) and (LATER). Source: PS Survey/ANAC (date of interview).

- PANDEMIC (LATER) is a dummy variable that takes the value of 1 for passengers interviewed in the later period of the COVID-19 pandemic, from January 2021 until the end of the sample in July 2021. No interviews were conducted between April 2020 and December 2020. The reference case for this binary variable is the year 2018. See PANDEMIC (PRE) and (EARLY). Source: PS Survey/ANAC (date of interview).

- PANDEMIC (PRE) is a dummy variable that takes the value of 1 for passengers interviewed in in the year preceding the onset of the COVID-19 pandemic, 2019. The reference case for this binary variable is the year 2018. See PANDEMIC (EARLY) and (LATER). Source: PS Survey/ANAC (date of interview).

- PRCONNECT is the proportion of passengers who will disembark from a connecting flight among the passengers on board the aircraft of the interviewed passenger's flight. Sources: PS Survey/ANAC (flight number) and ATSD-M/ANAC (number of connecting passengers and flight totals); constructed using authors' calculations.

- REDEYE is a dummy variable that takes the value 1 in case the interviewed passenger's flight departs between 11:00 pm on one day and 5:59 am on the following day. The reference case for this binary variable is the period between 6:00 am and 10:59 pm. Source: PS Survey/ANAC (flight number) and VRA/ANAC (flight time); constructed using authors' calculations and classifications.

- RUNWAYCONG is a proxy for the ratio of runway capacity usage at GRU Airport at the scheduled flight time of the interviewed passenger. To calculate this variable, all flights operated during the full hour of operation (takeoffs and landings) are considered, divided by the declared hourly capacity of using the airport's runways in the respective IATA-defined season (summer or winter). Sources: PS Survey/ANAC (flight number), VRA/ANAC (flight time, total number of landings and takeoffs performed at each date/time), and Airport Capacity (runway) Declaration of Guarulhos Airport – São Paulo/ANAC; constructed using authors' calculations and classifications.

- RUNWAYDIS is a proxy for runway disruption at GRU Airport. This is equal to the average proportion of flights that took off or landed late at the airport or were canceled over the full hour of the interviewed passenger's scheduled flight time. We define a delayed takeoff or landing as one that takes place 15 min after the scheduled time. Sources: PS Survey/ANAC (flight number) and VRA/ANAC (flight time, total scheduled and delayed takeoff and landing); constructed using authors' calculations and classifications.

- SCHLCOLL is a dummy variable that takes the value of 1 for the case of the interviewed passenger who declared himself as having up to the college level of education. See details in the SCHLELEM. Source: PS Survey/ANAC; constructed using authors' classifications.

- SCHLELEM is a dummy variable that takes the value of 1 if the interviewed passenger has declared himself as having up to the Elementary School level of education, either not initiated, unfinished, or finished. The original question in the survey questionnaire asked passengers to identify their education among the following alternatives: "(1) illiterate or unfinished Elementary School, (2) Elementary School, (3) Middle School, (4) High School, (5) unfinished college or university, and (6) college/university. As a reference case, this binary variable has passengers classified as having College/University, either finished or unfinished. See also SCHLHIGH and SCHLMIDD for more details. Source: PS Survey/ANAC; constructed using authors' classifications.

- SCHLHIGH is a dummy variable that takes the value of 1 if the interviewed passenger has declared himself as having a High School education level. See details in the SCHLELEM. Source: PS Survey/ANAC; constructed using authors' classifications.

- SCHLMIDD is a dummy variable that takes the value of 1 if the interviewed passenger has declared himself as having up to the Middle School level of education. See details in the SCHLELEM. Source: PS Survey/ANAC; constructed using authors' classifications.

- SHOPS measures the satisfaction of interviewed passengers with the quality and variety of retail concessions at the airport. A 5-point Likert scale was applied to the PS Survey/ANAC survey questionnaire to calculate the variable. The comments missing for this item in the survey were filled with a null value. The average responses regarding the quality and variety of services were recorded. The variable was rescaled to values between 0 and 1. The original question in the questionnaire was "*How do you evaluate the quality of stores/commercial establishments?*" and "*How do you evaluate the variety of stores/commercial establishments?*" Source: PS Survey/ANAC; constructed using authors' calculations.

- SHOPS (4/5 RATING) is a dummy variable that takes the value 1 if the passenger has rated the survey questionnaire items in the PS Survey/ANAC, which referred to the quality and variety of existing retail concessions at the airport with a minimum of 4 points out of 5 (represented by "Good" or "Very Good"). The questions in the original questionnaire were: "*How do you evaluate the quality of stores/commercial establishments?*" and "*How do you evaluate the variety of stores/commercial establishments?*" Source: PS Survey/ANAC; constructed using authors' calculations.

- SMALLTERM is a dummy variable that takes the value of 1 for passengers interviewed near boarding gates at Terminal 1 at GRU Airport. Terminal 1 is the smallest terminal at the airport and is currently mostly operated by Azul Airlines, generally with domestic flights. The reference case for this binary variable is Terminal 2 (older and with domestic hybrid operations and international operations in general with destinations in Latin America). See also INTNLTERM. Source: PS Survey/ANAC (place of interview); constructed using authors' classifications.

- TERMDEN is a proxy for passenger density per area available at each GRU Airport terminal during passenger interviews. This variable is computed from the total number of passengers moving through the respective airport terminal during the full hour when the interviewed passenger's flight was scheduled, divided by the estimated terminal area in square meters. This variable was multiplied by ten. Sources: PS Survey/ANAC (place and time of the interview), ATSD-M/ANAC (number of hourly passengers in the terminal on the respective day/time), Google Earth (terminal area); constructed using authors' calculations.

- TERMDIS is a proxy for the agglomeration of passengers on delayed flights waiting for the boarding call at each GRU Airport terminal during passenger interviews. This variable is equal to the proportion of passengers moving on flights with a delay of more than 15 min in relation to the total number of passengers moved in the respective airport terminal at the peak hour when the interviewed passenger's flight was scheduled. Sources: PS Survey/ANAC (place and time of the interview) and ATSD-M/ANAC (total hourly passengers and number of passengers delayed at the terminal on the respective day and time); constructed using authors' calculations.

- WEATHER (DST) is a proxy for unfavorable weather conditions at the destination airport at the scheduled flight time of the interviewed passenger. This variable is a dummy that takes the value 1 in any of the situations identified as alerts referring to low ceiling (below 600 feet), low visibility (below 1500 m), wind gust (above 27 knots in a wet runway situation and above 33 knots in a dry runway situation), and occurrences of hail and thunderstorms. Sources: PS Survey/ANAC (flight number) and IEM/Iowa State University (weather); constructed using authors' classifications.

- WEATHER (ORG) is a proxy for unfavorable weather conditions at the home airport (GRU) at the scheduled flight time of the interviewed passenger. See WEATHER (DST) for further details. Sources: PS Survey/ANAC (flight number) and IEM/Iowa State University (weather); constructed using authors' classifications.

- WIFI measures the interviewees' satisfaction with the Wi-Fi Internet service available at the airport. A 5-point Likert scale was applied to the PS Survey/ANAC survey questionnaire to calculate the variable. The comments missing for this item in the survey were filled with a null value. The variable was rescaled to values between 0 and 1. The original question in the questionnaire was, "*How do you evaluate the quality of wireless internet and other internet connections provided by the airport operator?*" Source: PS Survey/ANA; constructed using authors' calculations.

- WIFI (4/5 RATING) is a dummy variable that takes the value 1 if the passenger has rated the survey questionnaire items in the PS Survey/ANAC, which referred to the interviewees' satisfaction with the Wi-Fi Internet service available at the airport with a minimum of 4 points out of 5 (represented by "Good" or "Very Good"). The questions in the original questionnaire were: "*How do you evaluate the quality of wireless internet and other internet connections provided by the airport operator?*" Source: PS Survey/ANAC; constructed using authors' classifications.

**Table 5–Estimation results determinants of flight delays (DEL)**

| Variable | (1) |
|---|---|
| WEATHER (ORG) | 0.0290 |
| WEATHER (DST) | 0.1347*** |
| SMALLTERM | 1.5836*** |
| INTNLTERM | 0.7622*** |
| JETBRIDGE | -0.0269 |
| PRCONNECT | -0.1283 |
| LOADFAC | 0.2697*** |
| AIRCSIZE | -0.0019 |
| CARGO | 0.0552 |
| DISTANCE | -0.0296 |
| BUSYDAY | 0.2419*** |
| BUSYHOUR | 0.0615*** |
| SECINSPTIME | 0.1136 |
| RUNWAYCONG | 0.3785*** |
| RUNWAYDIS | 1.7516*** |
| CASCAD (DEP) | -0.0234 |
| CASCAD (ARR) | 2.0114*** |
| PANDEMIC (EARLY) | -0.0698 |
| PANDEMIC (LATER) | 0.3113*** |
| Estimator | RE-Probit |
| Log-likelihood | -5,167 |
| AIC Statistic | 10,376 |
| BIC Statistic | 10,533 |
| Nr Observations | 13,071 |

*Notes: Estimation results produced by ordered probit regression with robust standard errors. Variable selection performed by a previous procedure using the post-double-selection LASSO-based methodology (PDS-LASSO) of Belloni et al. (2012, 2014a,b). LASSO penalty loadings account for the clustering of airport terminal, and survey date. Flight date, airline, and destination control variables estimates omitted. Variables set as under LASSO penalization: DISSAT (all variables), and the proposed controls. P-value representations: \*\*\*p<0.01, \*\* p<0.05, \* p<0.10.*